\pdfoutput=1

\documentclass[12pt]{antonio}

\usepackage{graphicx}

\def\theequation{\thesection.\arabic{equation}}

\newcommand{\bea} {\begin{eqnarray*}}
\newcommand{\beq} {\begin{equation}}
\newcommand{\bey} {\begin{eqnarray}}
\newcommand{\eea} {\end{eqnarray*}}
\newcommand{\eeq} {\end{equation}}
\newcommand{\eey} {\end{eqnarray}}

\newcommand{\dif} {{\rm d}}
\newcommand{\dps} {\displaystyle}

\newcommand{\meio} {{1 \over 2}}

\newcommand{\nd} {\noindent}

\newcommand{\ptl} {\partial}

\newcommand{\vv} {\vspace{4mm}}

\newcommand{\dt} {\delta}

\newcommand{\gm} {\gamma}

\newcommand{\ta} {\theta}

\newcommand{\vf} {\varphi}

\newcommand{\calh} {{\cal H}}
\newcommand{\calt} {{\cal T}}

\newcommand{\nbb}{\mbox{\protect \boldmath $\nabla$}}

\begin{document}

\pagestyle{empty}

\vspace*{4cm}

\begin{center}

{\LARGE\bf
\begin{tabular}{c}
THEORY OF MULTIPOLE \\
\\
SOLUTIONS TO THE SOURCELESS \\
\\
GRAD-SHAFRANOV EQUATION \\
\\
IN PLASMA PHYSICS
\end{tabular}

}

\vspace{1.5cm}

{\Large\it by}

\vspace{1.5cm}

{\LARGE\bf A. FERREIRA}

\vfill


\end{center}

\newpage

\vspace*{2cm}

\begin{center}

{\LARGE\bf
\begin{tabular}{c}
Theory of multipole solutions \\
to the sourceless \\
Grad-Shafranov equation \\
in plasma physics
\end{tabular}

}

\vspace{1.5cm}

{\Large\bf A. Ferreira}

\vv

\textit{R. Goi\'as, 1021, Jardim Santa Cruz}

\textit{18700-140 \ Avar\'e, S\~ao Paulo, Brazil}

\end{center}

\vspace{3cm}

\centerline{{\large\bf Abstract}}

\vv

\nd \textit{ The rules to write out any one of the linearly
independent functions belonging to the infinite set of those in
polynomial form that satisfy the sourceless Grad-Shafranov equation
as stated in the toroidal-polar coordinate system are established.
It is found that a polynomial solution even in the poloidal angle is
given by the product of an integral power of the radial coordinate
variable by a complete polynomial of equal degree in this same
variable with angular-dependent coefficient functions that are
linear combinations of a finite number of Chebyshev polynomials in
the cosine of the poloidal angle, the numerical coefficients of
these being expressed in terms of the binomial numbers of Pascal's
arithmetic triangle. Tables of the ten polynomial solutions of the
lowest degrees are provided in variables of the toroidal-polar and
of the cylindrical coordinate systems.}

\vfill


\newpage

\pagestyle{plain} \setcounter{page}{1} \baselineskip=8.5mm

\centerline{{\bf I. INTRODUCTION}}

\setcounter{section}{1}

\vv

In the literature of Plasma Physics devoted to the equilibrium of
the toroidal pinch, multipole solutions are generally understood as
solutions to the Grad-Shafra\-nov equation with no source terms as
represented by the gradient of the plasma pressure and the gradient
of half the squared toroidal field function (also called poloidal
current function) in flux space, some examples of which have been
known since the early days of thermonuclear fusion research
\cite{um}, \cite{dois}. In the present paper we shall designate by
\textit{multipole fields} the magnetic fields in vacuum that are
invariant under rotation about a fixed axis in space and whose field
lines are entirely contained in the planes passing by the symmetry
axis; the fluxes of such fields coincide then with the multipole
solutions in the sense this expression is costumarily employed in
Plasma Physics. \footnote{It should be noted that an azimuthal
magnetic field that falls with the inverse of the distance away from
the axis of rotational symmetry and that the presence of a
conducting fluid with a uniform pressure distribution in space still
do not provide the Grad-Shafranov equation with a source term. We
shall exclude such magnetic field and material medium from our
definition of a multipole field.}

Multipole solutions have found theoretical application in the
solution of the free boundary problem of tokamak plasmas, and, by
extension, in the calculation of external magnetic field
configurations required by such confining devices and design of the
system of coils intended to generate them \cite{dois}. As it is our
aim to show in the article that follows this in the current issue of
this journal \cite{quatro}, the utility of the multipole solutions
goes beyond this scope, standing their use on the basis of a method
of solution to the Grad-Shafranov equation for which the sources are
constant in flux space. This multiplicity of utilizations then
justifies the study of the solutions to the partial differential
equation whose derivation we pass now to outline \cite{cinco}.

For frame of reference we choose the cylindrical coordinate system
as illustrated in Fig. 1, which is, among all rotational coordinate
systems, the one with the simplest metrical properties\footnote{The
cyclic order of the unity vectors in the cylindrical system
represented in Fig. 1, with the left hand rule being presumed, is
$(\vec e_R, \vec e_\phi, \vec k)$. Since this order is the same as
that of the unity vectors of same names in the conventional
right-handed cylindrical coordinate system, the results for all
vector operations in both systems have equal forms. In particular,
the terms in the expressions for the curl of a vector $(B_R, B_\phi,
B_z)$ have the same respective signals in one and the other
systems.}.

\begin{figure}
\begin{center}

{\unitlength=1mm
\begin{picture}(120,92)
\put(0,0){\includegraphics{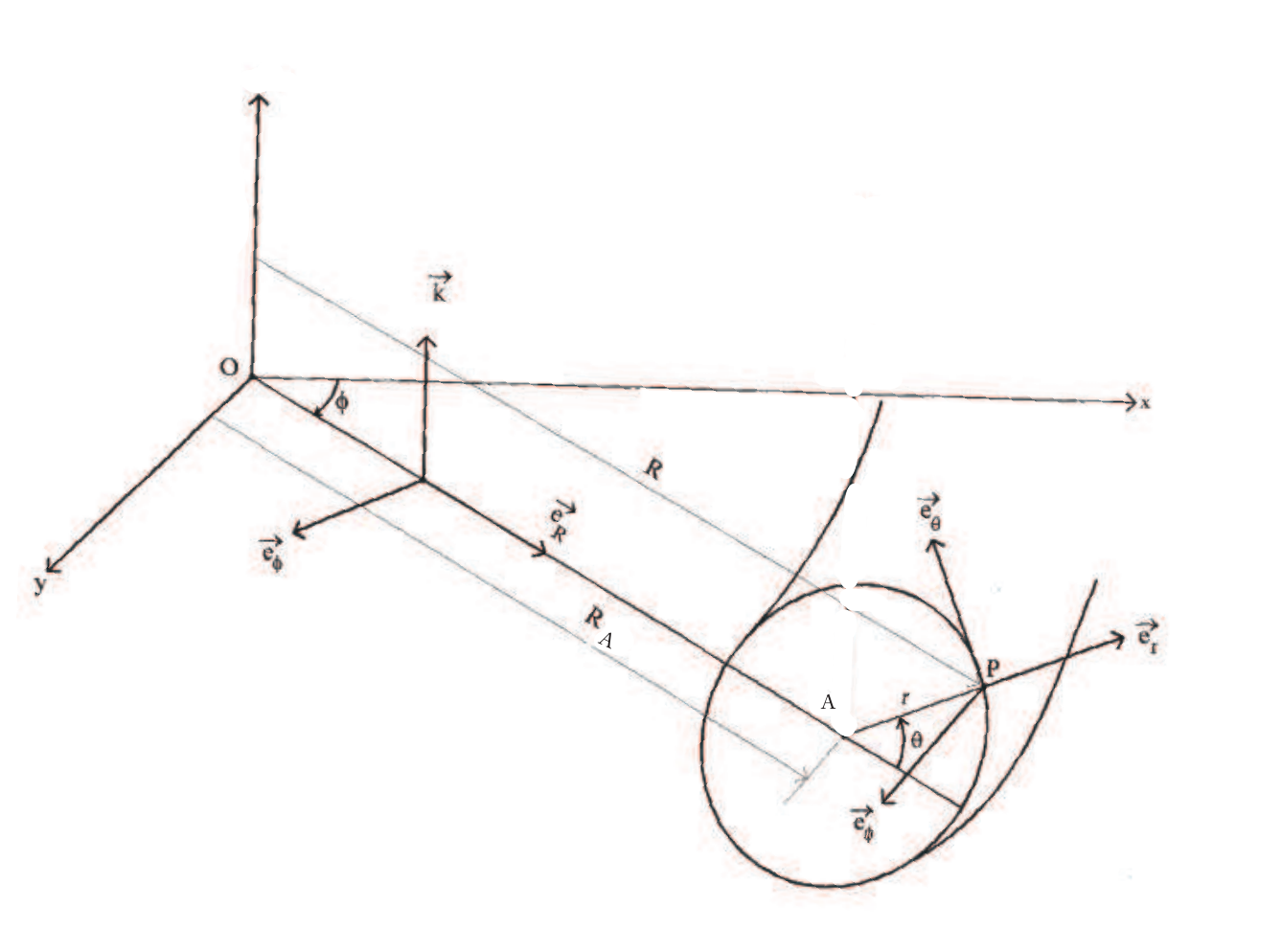}}
\put(30,90){{\footnotesize z $\equiv$ axis of rotational symmetry}}
\end{picture}
}

\vv\vv\vv

\begin{tabular}{lp{12cm}}
{\bf FIG. 1} & The left-handed Cartesian coordinate system
$(x,y,z)$, the left-handed cylindrical coordinate system $(R, \phi,
z)$ and the right-handed toroidal-polar coordinate system $(r, \ta,
\phi)$.
\end{tabular}

\end{center}

\vv

\end{figure}

The multipole fields satisfy the equations: \beq \nbb.\vec B = 0
\label{eq1.1} \eeq and \beq \nbb\times\vec B = 0\ . \label{eq1.2}
\eeq

The solenoidal law, which in the coordinate system of our choice
takes the form: \beq \frac{1}{R}\frac{\ptl}{\ptl R}(RB_R) +
\frac{\ptl B_z}{\ptl z} = 0 \ , \label{1.3nova} \eeq is
automatically satisfied if the radial and the axial components of
the magnetic field are written in terms of a scalar ``stream
function'' $\Psi$ as: \bey B_R &=&
-\frac{1}{R} \frac{\ptl\Psi}{\ptl z}\ , \label{eq1.3} \\
B_z &=& \frac{1}{R} \frac{\ptl\Psi}{\ptl R}\ . \label{eq1.4} \eey

For a rotationally symmetric vector field with components lying on
meridian planes the curl admits a nonnull projection only on the
azimuthal $\phi$-direction, which, in the left-handed cylindrical
coordinate system $(R, \phi, z)$ represented in Fig. 1, is expressed
as: \beq (\nbb \times \vec B)_\phi = \frac{\ptl B_R}{\ptl z} -
\frac{\ptl B_z}{\ptl R}\ . \label{eq1.5} \eeq

Replacing $B_R$ and $B_z$ as given by Eqs. (\ref{eq1.3}) and
(\ref{eq1.4}) in Eq. (\ref{eq1.5}), we can cast Amp\`ere's law
into the form: \beq R^2 \nbb.\left(\frac{1}{R^2}\nbb\Psi\right) =
0 \ , \label{eq1.6} \eeq which is, in vector notation, the
equation governing the multipole fields. The stream function
$\Psi$ can be shown \cite{cinco} to be physically the flux
associated with the meridian (or poloidal) magnetic field divided
by $2\pi$.

Equation (\ref{eq1.6}), which we shall call the sourceless
Grad-Shafranov equation or the multipole equation, is a partial
differential equation of the elliptic type which, referring to the
two coordinate systems most widely used in equilibrium studies, the
cylindrical and the toroidal-polar (see Fig. 1), admits of
infinitely many solutions in the form of polynomials in one
coordinate variable with the meaning of length ($z$ or $r$), the
coefficients of which are themselves polynomials, if the first of
those two systems is used, also in a linear coordinate variable
($R$), or, if the second one is used, in the cosine of the angular
coordinate ($\ta$). With reference to an equatorial plane in space
defined by the coordinate $z = 0$, these solutions can be grouped in
even ones and in odd ones, and, inside each parity group, linearly
independent solutions can be singled out according to the degree of
the polynomial they are in a linear coordinate variable (say). From
a mathematical viewpoint it is these independent solutions that we
shall recognize as the multipole solutions.

It is the object of the present paper to derive the infinite set of
the even multipole solutions to the sourceless Grad-Shafranov
equation according to the foregoing definition here given to them.
Of the two previously cited coordinate systems, our choice will fall
upon the toroidal-polar as the ground for the analytical work, since
the developments in this system lead naturally to the representation
of the angular dependence of the solutions in terms of orthogonal
polynomials that have simple and well known properties.

We shall commence by assigning the multipole solutions the general
form of polynomials in the radial coordinate variable with
coefficients that are unknown functions of the cosine of the angular
coordinate variable.\footnote{This is true for solutions of both
parities; the odd ones are made so by effect of an overall
multiplying sine factor.} From the construction of a few examples of
the lowest degrees, the precise form of polynomial dependence the
linearly independent solutions must bear on the radial coordinate
variable is inferred. The ensuing problem to tackle is that of the
dependence of the solutions on the angular variable, and accordingly
it takes the guise, no longer of a partial differential equation in
two variables, but that of ordinary differential equations in the
angular one for the coefficient functions of the polynomials whose
dependence on the radial coordinate variable has already been
established. It is at this stage of the analysis that the
representation of the angular dependences of the multipole solutions
in terms of orthogonal polynomials appears forcefully as the
analytical recourse to be adopted since they are advanced by the
very form of the differential equations governing those dependences.
Two families of orthogonal polynomials present themselves as natural
candidates to play the role of basis for the (finite) expansion of
the angular-dependent coefficients: that of the Chebyshev
polynomials and that of the associated Legendre polynomials of order
unity, of which we have given preference to the former, on account
both of the simpler defining properties of its members and of the
form we should lend to the solution as suggested by the isolated
examples worked out.

The finite set of Chebyshev polynomials that are to enter a
multipole solution of a given degree in the radial variable is found
by a blend of general arguments and induction. The next task in the
progression is then to determine the numerical coefficients by which
they must be multiplied in a combination that satisfies the ordinary
differential equations for the coefficient functions of the assumed
polynomial solution.

At this level the problem will appear as transmuted from that of
differential equations, partial and ordinary, into that of a
restricted set of difference equations for the numerical
coefficients of nonseparable solutions in two variables. Three are
these difference equations, two of them ordinary and one partial,
the solutions of the ordinary ones serving as boundary conditions
to the partial one.

Besides the recurrent construction of the set of numerical
coefficients entering the multipole solution of a particular degree
they allow for, we show that these difference equations can be given
solutions under the form of general expressions by which capability
the step-by-step procedure of evaluation of the coefficients can be
circumvented. Moreover, since the end values are \textit{ipso facto}
contained in the solution to the partial difference equation, this
one comprises also the solutions to the ordinary difference
equations and a single expression which attends to all the
coefficients belonging to any and all multipole solutions can be
written.

Up to the present moment uses seem not to have been ever made of the
odd multipole solutions, which however would find application in
systems of confinement that would not exhibit up-down symmetry with
respect to the equatorial plane of the torus. In the present paper
we shall not pursue the full characterization of such odd solutions,
limiting ourselves to showing how they arise along a process of
systematic construction of polynomial solutions to the multipole
equation.

\vv\vv

\begin{center}
{\bf II. THE RADIAL DEPENDENCE OF THE LINEARLY \\
INDEPENDENT MULTIPOLE SOLUTIONS}
\end{center}

\setcounter{section}{2} \setcounter{equation}{0}

\vv

As stated in Section I, we shall adopt the toroidal-polar coordinate
system in which to express the multipole equation and to work out
its solutions. This system, which is pictorially represented in Fig.
1, is formed by rotating the usual polar system of the plane about a
fixed axis in space, which we identify with the $z$-axis. In the
planar system, the polar axis is placed perpendicularly to the axis
of rotation; the pole $A$ is located on the polar axis at a distance
$R_A$ from that of rotational symmetry. As the plane of the system
is revolved, the polar axis generates a plane, which we refer to as
the equatorial plane, perpendicular to the $z$-axis. We shall call
meridian plane any plane containing the $z$-axis. The coordinates of
a point $P$ belonging to the meridian plane determined by the
$z$-axis and a given position of the polar axis are then the radius
$r$, which measures the distance from the pole $A$ to $P$, the polar
angle $\ta$, the angle between the positive direction of the polar
axis (the direction of growing distances from the $z$-axis) and the
radius, and the azimuthal angle $\phi$, the angle of rotation given
to a reference plane passing by the axis of rotation to bring that
plane from an arbitrarily chosen conventional position, to which we
assign the azimuthal angle $\phi = 0$, to coinciding with the
meridian plane containing the $z$-axis, the polar axis and the point
$P$, such that $(r,\ta,\phi)$ form a positive set according to the
right-hand rule. In this coordinate system Eq. (\ref{eq1.6}) takes
the form: \beq x^2(1 + x\mu)\frac{\ptl^2\psi}{\ptl x^2} + (1 +
x\mu)(1-\mu^2)\frac{\ptl^2\psi}{\ptl\mu^2} + x\frac{\ptl\psi}{\ptl
x} - (x + \mu)\frac{\ptl\psi}{\ptl\mu} = 0 \ , \label{eq2.1} \eeq
where $\mu \equiv \cos\ta$, \beq x \equiv \frac{r}{R_A}
\label{eq2.2} \eeq is the radial coordinate normalized to $R_A$ and
$\psi$ is the flux function $\Psi$ normalized to an arbitrary flux
$\Psi_0$. From now on it is Eq. (\ref{eq2.1}) that we shall refer to
as the sourceless Grad-Shafranov equation or the multipole equation.

We shall recognize that Eq. (\ref{eq2.1}) has infinitely many
linearly independent solutions of polynomial form in the variable
$x$ with coefficients that are polynomials in the variable $\mu$.
Considered as functions of the angle $\ta$, these independent
solutions can be grouped into two sets, of the even ones and of the
odd ones; we shall designate the solutions belonging to the first
set by the Greek letter $\vf$ and those belonging to the second set
by the Greek letter $\gm$.

Since in Eq. (\ref{eq2.1}) only derivatives of the unknown
function appear, a constant is a solution, and we write: \beq
\vf^{(0)}(x, \mu) = 1\ , \label{eq2.3} \eeq which we refer to as
the even multipole solution of order zero.

A polynomial solution of the first degree in $x$ writes in general
as: \beq \psi(x, \mu) = w_0(\mu) + w_1(\mu)x\ . \label{eq2.4} \eeq

To determine the forms of the functions $w_0(\mu)$ and $w_1(\mu)$ we
substitute this expression in the multipole equation. Collecting
terms of equal power in $x$ and equating the resulting coefficient
of each power to zero, we obtain from the zeroth power: \beq
(1-\mu^2)\frac{\dif^2w_0}{\dif\mu^2} - \mu\frac{\dif w_0}{\dif\mu} =
0\ . \label{eq2.5} \eeq

The solution of this differential equation is: \beq w_0(\mu) = a_0 +
c_0\arccos\mu\ , \label{eq2.6} \eeq where $a_0$ and $c_0$ are two
arbitrary constants. Since our interest is restricted to solutions
that are periodic in the variable $\ta = \arccos\mu$, we demand
$c_0$ to be zero, and $w_0(\mu)$ thus reduces to:
$$
w_0(\mu) = a_0\ . \eqno(2.6a)
$$

We next consider the equation resulting from putting the coefficient
of the first power in $x$ equal to zero. We have: \bey
(1-\mu^2)\frac{\dif^2w_1}{\dif\mu^2} - \mu\frac{\dif w_1}{\dif\mu} +
w_1 &=& -\mu(1-\mu^2)\frac{\dif^2 w_0}{\dif\mu^2} + \frac{\dif
w_0}{\dif\mu} \nonumber \\
&=& 0\ , \label{eq2.7} \eey the second equality coming from the
use of equation (2.6a) for $w_0(\mu)$ in the right hand side of
the first equality. This equation can be recognized as an instance
of Chebyshev's differential equation \cite{sete}, whose solution
can be written as: \beq w_1(\mu) = a_{11}T_1(\mu) +
b_{10}\sqrt{1-\mu^2} U_0(\mu)\ , \label{eq2.8} \eeq where \beq
T_1(\mu) = \mu \label{eq2.9} \eeq is the Chebyshev polynomial type
I of the first order, \beq U_0(\mu) = 1 \label{eq2.10} \eeq is the
Chebyshev polynomial type II of the zeroth order, and $a_{11}$ and
$b_{10}$ are arbitrary constants.

The terms now remaining in the equation obtained by substituting Eq.
(\ref{eq2.4}) in the multipole equation are all of the second power
in $x$. At this last level in the succession of equations for the
$\mu$-dependence of the assumed polynomial solution we meet a
condition again regarding $w_1(\mu)$, to know: \beq
\mu(1-\mu^2)\frac{\dif^2 w_1}{\dif\mu^2} - \frac{\dif w_1}{\dif\mu}
= 0\ . \label{eq2.11} \eeq The solution of this differential
equation can be easily obtained and is: \beq w_1(\mu) = c_1 +
d_1\sqrt{1-\mu^2} \ , \label{eq2.12} \eeq where $c_1$ and $d_1$ are
arbitrary constants.

To make equal the two expressions we have for $w_1(\mu)$ we require
that \beq a_{11} = 0 \label{eq2.13} \eeq in Eq. (\ref{eq2.8}) and
that \bey c_1
&=& 0 \ , \label{eq2.14} \\
d_1 &=& b_{10} \label{eq2.15} \eey in Eq. (\ref{eq2.12}). Note that
this choice for the free constants $a_{11}$ and $c_1$ suppresses the
terms that are even in $\ta$ in each of the two distinct expressions
for $w_1(\mu)$.

With $w_0(\mu)$ and $w_1(\mu)$ so determined, the expression that
results for the solution of the form proposed in Eq. (\ref{eq2.4})
is: \beq \psi(x, \mu) = a_0 + b_{10}x\sqrt{1-\mu^2} U_0(\mu)\ .
\label{eq2.16} \eeq

The constant term $a_0$ is just a multiple of the multipole
solution $\vf^{(0)}(x, \mu)$, and the new, independent solution
brought about by Eq. (\ref{eq2.16}) is: \beq \gm^{(0)}(x, \mu) =
x\sqrt{1-\mu^2} U_0(\mu)\ , \label{eq2.17} \eeq where we have
adopted, to designate it, the notation appropriate to an odd
multipole solution in the angular variable, which it is. We shall
call this the odd multipole solution of the zeroth order.

We next consider a polynomial solution of the second degree in $x$
for the multipole equation: \beq \psi(x,\mu) = w_0(\mu) + w_1(\mu)x
+ w_2(\mu)x^2\ . \label{eq2.18} \eeq

The procedure of substitution in Eq. (\ref{eq2.1}) and of equating
the coefficients of the powers of $x$ to zero leads again, from the
lowest power, to Eq. (\ref{eq2.5}) for $w_0(\mu)$, which is solved
by Eq. (2.6a) as before. From the equality of the term in $x$ to
zero we obtain Eq. (\ref{eq2.7}) for $w_1(\mu)$, the solution of
which is given by Eq. (\ref{eq2.8}). Since a condition like that of
Eq. (\ref{eq2.11}) is no longer found when we go to the next power
of $x$, we keep both terms in Eq. (\ref{eq2.8}) (that is to say, we
take $a_{11} \ne 0$).

The term in $x^2$ in the multipole equation gives rise to the
equation: \bey (1-\mu^2)\frac{\dif^2w_2}{\dif\mu^2} -
\mu\frac{\dif w_2}{\dif\mu} + 4w_2 &=&
-\mu(1-\mu^2)\frac{\dif^2w_1}{\dif\mu^2} + \frac{\dif
w_1}{\dif\mu} \nonumber \\
&=& a_{11} \ , \label{eq2.19} \eey where the second equality
follows from the use of Eq. (\ref{eq2.8}) for $w_1(\mu)$ in the
right hand side of the first equality. We are again in the
presence of a Chebyshev equation, this time inhomogeneous, the
complete solution of which is: \beq w_2(\mu) = a_{22}T_2(\mu) +
b_{21}\sqrt{1 - \mu^2} U_1(\mu) + \frac{a_{11}}{4} \ ,
\label{eq2.20} \eeq where $T_2(\mu)$ and $U_1(\mu)$ are
respectively the Chebyshev polynomial type I of the second order,
defined by \beq T_2(\mu) = 2\mu^2 - 1\ , \label{eq2.21} \eeq and
the Chebyshev polynomial type II of the first order, defined by
\beq U_1(\mu) = 2\mu \ , \label{eq2.22} \eeq and $a_{22}$ and
$b_{21}$ are two arbitrary constants.

The constraint that results from equating the term in $x^3$ in the
multipole equation to zero closes the set, and, the same as in the
previous case of a polynomial solution of the first degree in $x$,
takes on the form of a homogeneous differential equation for the
$\mu$-dependent coefficient of the term of the highest power in $x$
in the proposed solution, in this way imposing a demand on
$w_2(\mu)$ of its own. We have: \beq
\mu(1-\mu^2)\frac{\dif^2w_2}{\dif\mu^2} - \frac{\dif w_2}{\dif\mu} +
2\mu w_2 = 0\ . \label{eq2.23} \eeq

In general, for an assumed polynomial solution of any degree, by
following the two opposite leads in the chain of equations for the
coefficient functions, one corresponding to the equation associated
with the lowest power, and the other corresponding to the equation
associated with the highest power of $x$ in the multipole equation,
we are able to generate two parallel sets of solutions for the
$\mu$-dependent functions. All the consistency of the method of
construction of polynomial solutions and in fact the very existence
of such solutions hinge on the possibility of making equal these two
sets by appropriately choosing the arbitrary constants appearing in
both.

A possible manner of bringing $w_2(\mu)$, as given by Eq.
(\ref{eq2.20}), to satisfy Eq. (\ref{eq2.23}) is simply by
substituting the former into the latter and then choosing for the
constants $a_{11}$, $a_{22}$ and $b_{21}$ the values that solve the
ensuing system of algebraic equations. An alternative manner
consists in comparing the solution of Eq. (\ref{eq2.23}) with the
expression for $w_2(\mu)$ stated in Eq. (\ref{eq2.20}). Adopting
this second one, we write down the solution of Eq. (\ref{eq2.23}):
\beq w_2(\mu) = c_2\mu^2 + d_2\left[\sqrt{1-\mu^2} +
\frac{\mu^2}{2}\ln\left(\frac{1+\sqrt{1-\mu^2}}{1 -
\sqrt{1-\mu^2}}\right)\right]\ , \label{eq2.24} \eeq where $c_2$ and
$d_2$ are constants, and then, by equating it to the solution for
$w_2(\mu)$ as given by Eq. (\ref{eq2.20}), we are led to determine
the constants which appear in both as: \bey a_{22} &=&
\frac{a_{11}}{4} \ , \label{eq2.25} \\
b_{21} &=& d_2 = 0\ , \label{eq2.26} \\
c_2 &=& 2a_{22} \ . \label{eq2.26lin} \eey

Note that these choices for $b_{21}$ and $d_2$ have the effect of
suppressing the odd functions in $\ta$ that enter the expressions of
the two solutions obtained for $w_2(\mu)$.

With all these results at hand we are in position to state the
polynomial solution of the second degree in $x$ for the multipole
equation as: \beq \psi(x, \mu) = a_0 +
b_{10}x\sqrt{1-\mu^2}U_0(\mu) + a_{11}\left\{T_1(\mu)x +
\left[\frac{1}{4}T_0(\mu) + \frac{1}{4}T_2(\mu)\right]x^2\right\}.
\label{eq2.27} \eeq

Now, the first two terms on the right hand side are just a
combination of the two previously found multipole solutions
$\vf^{(0)}(x, \mu)$ and $\gm^{(0)}(x, \mu)$, and we recognize the
third, new term, as a multiple of the even multipole solution of
the first order, which we shall denote by $\vf^{(1)}(x, \mu)$:
\beq \vf^{(1)}(x,\mu) = T_1(\mu)x + \left[\frac{1}{4}T_0(\mu) +
\frac{1}{4}T_2(\mu)\right]x^2 \ . \label{eq2.28} \eeq

If we next consider a polynomial solution of the third degree in
$x$, by following the same steps of the procedure we employed to
derive the even and odd multipole solutions of lower degrees, we
arrive at the odd multipole solution of the first order, which is:
\beq \gm^{(1)}(x,\mu) = x\sqrt{1-\mu^2}\left\{U_1(\mu)x +
\left[\frac{1}{4}U_0(\mu) + \frac{1}{4}U_2(\mu)\right]x^2\right\}
\ , \label{eq2.29} \eeq where \beq U_2(\mu) = 4\mu^2 - 1
\label{eq2.30} \eeq is the Chebyshev polynomial type II of the
second order.

The examination of these and of higher order multipole solutions in
this way obtained shows that they have definite parity, being either
even or odd in the polar angle $\ta$, the general form of the even
ones of order $n$ being: \beq \vf^{(n)}(x,\mu) =
\sum_{k=n}^{2n}f_k^{(n)}(\mu)x^k\ , \label{eq2.31} \eeq and that of
the odd ones of order $n$ being: \beq \gm^{(n)}(x,\mu) =
x\sqrt{1-\mu^2}\sum_{k=n}^{2n}g_k^{(n)}(\mu)x^k\ , \label{eq2.32}
\eeq where the angular functions $f_k^{(n)}(\mu)$ and
$g_k^{(n)}(\mu)$ should be expected to express themselves naturally
as combinations of Chebyshev polynomials type I and type II
respectively. The problem then reduces to finding the coefficients
of these combinations given the order $n$ of the multipole solution.
In this paper we shall treat only the case of the solutions that are
even with respect to the polar angle $\ta = \arccos\mu$.

\vv\vv

\begin{center}
{\bf III. THE DIFFERENTIAL EQUATIONS FOR THE ANGULAR
\\ DEPENDENCES OF THE MULTIPOLE SOLUTIONS; \\ THE REPRESENTATION OF
THE SOLUTIONS AS \\ COMBINATIONS OF CHEBYSHEV POLYNOMIALS AND
\\ THE RECURRENCE RELATIONS FOR THEIR COEFFICIENTS}
\end{center}

\setcounter{section}{3}  \setcounter{equation}{0}

\vv

We start by introducing Eq. (\ref{eq2.31}) in Eq. (\ref{eq2.1}).
Collecting terms of equal powers of $x$ and equating the resulting
coefficient for each power to zero, we obtain three differential
equations for the angular functions $f_k^{(n)}(\mu)$ according to
the value of the exponent of $x$. From the term in $x^n$ we obtain
an equation for $f_n^{(n)}(\mu)$, which is: \beq
(1-\mu^2)\frac{\dif^2f_n^{(n)}}{\dif\mu^2} - \mu\frac{\dif
f_n^{(n)}}{\dif\mu} + n^2f_n^{(n)} = 0\ , \label{eq3.1} \eeq and,
from the term in $x^{2n+1}$, an equation for $f_{2n}^{(n)}(\mu)$,
namely: \beq \mu(1-\mu^2)\frac{\dif^2f_{2n}^{(n)}}{\dif\mu^2} -
\frac{\dif f_{2n}^{(n)}}{\dif\mu} + 2n(2n-1)\mu f_{2n}^{(n)} = 0\ ,
\label{eq3.2} \eeq both of which are homogeneous. The coefficient of
$x^\ell$, for $\ell = n+1, n+2, \ldots, 2n$, gives rise to a
differential relation involving two angular functions of subscripts
differing by unity, which can be written as: \bey
&&\hspace{-6mm}(1-\mu^2)\frac{\dif^2f_\ell^{(n)}}{\dif\mu^2} -
\mu\frac{\dif f_\ell^{(n)}}{\dif\mu} + \ell^2f_\ell =
-\mu(1-\mu^2)\frac{\dif^2f_{\ell - 1}^{(n)}}{\dif\mu^2} + \frac{\dif
f_{\ell -1}^{(n)}}{\dif\mu} - (\ell - 1)(\ell - 2)\mu
f_{\ell-1}^{(n)} \nonumber \\
&&\hspace{-6mm}(\ell = n+1, n+2,\ldots, 2n). \label{eq3.3} \eey

We now proceed by considering each of these equations.

\vv

\nd \textbf{(1) The equation for $f_n^{(n)}(\mu)$}

Equation (\ref{eq3.1}) can be immediately recognized as being
Chebyshev's equation, whose even solution is the Chebyshev
polynomial type I of order $n$, $T_n(\mu)$ \cite{sete}. We write
thus: \beq f_n^{(n)}(\mu) = A_{n,n}^{(n)}T_n(\mu) \ (n = 0, 1,2,
3, \ldots), \label{eq3.4} \eeq where $A_{n,n}^{(n)}$ is a
constant.

\vv

\nd \textbf{(2) The homogeneous equation for $f_{2n}^{(n)}(\mu)$}

Although our ultimate aim is to express $f_{2n}^{(n)}(\mu)$ as a
combination of Chebyshev polynomials, it is nonetheless of interest
to consider also other representations which bear a more direct
reference to the general pattern the differential equation we found
to govern this function obeys. Indeed, the even solution in $\ta$ of
Eq. (\ref{eq3.2}) finds a concise representation as: \beq
f_{2n}^{(n)}(\mu) = \mu P_{2n-1}^1(\sqrt{1-\mu^2})\ , \label{eq3.5}
\eeq where $P_{2n-1}^1$ is the associated Legendre function of
degree $2n-1$ and order 1 of the first kind \cite{sete}. (The second
independent solution to Eq. (\ref{eq3.2}), which writes as:
\[
\mu Q_{2n-1}^1(\sqrt{1-\mu^2}),
\]
$Q_{2n-1}^1$ being the associated Legendre function of degree $2n-1$
and order 1 of the second kind, though it is regular in the whole
interval of variation of $\mu$, $-1 \le \mu \le 1$, is however odd
in the angle $\ta$ and must thence be rejected.) Another
representation of $f_{2n}^{(n)}(\mu)$ that we shall find to be
useful is obtained by noting that, upon the transformation: \beq
\mu^2 = t\ , \label{eq3.6} \eeq Eq. (\ref{eq3.2}) becomes: \beq
t(1-t)\frac{\dif^2f_{2n}^{(n)}}{\dif t^2} - \frac{t}{2}\frac{\dif
f_{2n}^{(n)}}{\dif t} + n\left(n - \meio\right)f_{2n}^{(n)} = 0\ ,
\label{eq3.7} \eeq which can be recognized as a form of the
hypergeometric equation \cite{sete}. The solution of interest for
$f_{2n}^{(n)}(\mu)$ can be written down immediately as: \beq
f_{2n}^{(n)}(\mu) = \mu^2\,_2F_1\left(-n+1, n+\meio; 2;
\mu^2\right)\ . \label{eq3.8} \eeq Since the first argument of the
hypergeometric function $_2F_1$ vanishes for $n=1$ and is a negative
integer for $n=2, 3, 4, \ldots$, this solution is actually a
polynomial of degree $n$ in $\mu^2$, for which the constant term is
missing. (To obtain a convenient representation of the second
solution to Eq. (\ref{eq3.2}), the transformation of the independent
variable $\mu^2 = 1 - v$ proves to be more advantageous than that of
Eq. (\ref{eq3.6}). By the sole consideration of the form into which
the equation is converted when the free variable passes to be $v$,
it becomes apparent that one of the independent solution for
$f_{2n}^{(n)}(\mu)$ is \cite{oito}:
\[
f_{2n}^{(n)}(\mu) = \sqrt{1-\mu^2}\,_2F_1\left(\meio - n, n;
\frac{3}{2};1-\mu^2\right)\ .
\]
Because of the sine term multiplying the hypergeometric function,
this solution is odd in the variable $\ta$ and therefore to be
abandoned.)

We now turn ourselves to our main objective with regard to the
angular function $f_{2n}^{(n)}(\mu)$, which is to find for it a
representation in terms of the Chebyshev polynomials. For this
purpose we shall take as starting point the differential equation
for $f_{2n}^{(n)}(\mu)$ itself rather than the explicit
definitions we have found by solving it. Equation (\ref{eq3.2})
can be concisely written as: \beq \calh_{2n}\{f_{2n}^{(n)}(\mu)\}
= 0\ , \label{eq3.9} \eeq where we have made use of the notation:
\beq \calh_\ell \equiv \mu(1-\mu^2)\frac{\dif^2}{\dif\mu^2} -
\frac{\dif}{\dif\mu} + \ell(\ell - 1)\mu\ , \label{eq3.10} \eeq
$\ell$ being an integer. For the attainment of our aim we have to
know which are the effects of the differential operator above
introduced as it acts on the Chebyshev polynomials in general.

By having recourse to the frequently quoted relations for the
Chebyshev polynomials \cite{sete}, \cite{nove}: \bey
(1-\mu^2)\frac{\dif^2T_m}{\dif\mu^2} &=& \mu\frac{\dif
T_m}{\dif\mu} - m^2T_m(\mu)\ , \label{eq3.11} \\
(1-\mu^2)\frac{\dif T_m}{\dif\mu} &=&
\frac{m}{2}\left[T_{m-1}(\mu) - T_{m+1}(\mu)\right]\ ,
\label{eq3.12} \\
2\mu T_m(\mu) &=& T_{m+1}(\mu) + T_{m-1}(\mu)\ , \label{eq3.13}
\eey the following properties of the operator $\calh_\ell$ can be
easily established: \bey \hspace{-7mm}\calh_\ell\{T_0(\mu)\} &=&
\ell(\ell-1)T_1(\mu)\ , \label{eq3.14} \\
\hspace{-7mm}\calh_\ell\{T_m(\mu)\} &=& \meio(\ell \!+\! m)(\ell
\!-\! m\!-\! 1)T_{m-1}(\mu) \!+\! \meio(\ell \!- \!m)(\ell \!+\!
m\!-\! 1)T_{m+1}(\mu) \label{eq3.15} \eey ($m = 1, 2, 3, \ldots;
\ell = 1, 2, \ldots, m$).

Next, from the considerations on the form of $f_{2n}^{(n)}(\mu)$
brought forth in the sequel of Eq. (\ref{eq3.8}), it is apparent
that this function can be suitably expressed as a finite combination
of Chebyshev polynomials, restricted these to the ones of the orders
smaller than and equal to $2n$, and even, to know: \beq
f_{2n}^{(n)}(\mu) = \sum_{\ell=0}^n A_{2n,2\ell}^{(n)}T_{2\ell}(\mu)
\ , \label{eq3.16} \eeq where the $A_{2n,2\ell}^{(n)}$'s are
constants.

If now we insert Eq. (\ref{eq3.16}) in Eq. (\ref{eq3.9}), by making
use of the formulae stated in Eqs. (\ref{eq3.14}) and
(\ref{eq3.15}), we are able to transform the differential equation
for $f_{2n}^{(n)}(\mu)$ into a null identity for a combination of
Chebyshev polynomials of odd orders, whose coefficients are
combinations two by two respectively of the constants
$A_{2n,2\ell}^{(n)}$'s. By appeal to the orthogonality property of
the Chebyshev polynomials it follows that the coefficient of each
polynomial in the identity must in its turn be null, and this gives
a recursive pair of formulae for the constants. From the coefficient
of $T_1(\mu)$ we get: \beq A_{2n,2}^{(n)} =
-2\frac{n(2n-1)}{(n+1)(2n-3)}A_{2n,0}^{(n)}\ , \label{eq3.17} \eeq
and from the coefficient of $T_{2\ell +1}(\mu)$ we get: \beq
A_{2n,2\ell + 2}^{(n)} = -\frac{(n-\ell)(2n+2\ell -1)}{(n+\ell+1)
(2n-2\ell-3)}A_{2n,2\ell}^{(n)} \ \ \ (\ell = 1, 2, \ldots, n-1).
\label{eq3.18} \eeq

Once $A_{2n,0}^{(n)}$ is specified, Eqs. (\ref{eq3.17}) and
(\ref{eq3.18}) provide us with the means to evaluate the remaining
$n$ coefficients $A_{2n,2\ell}^{(n)}$ that enter the making up of
$f_{2n}^{(n)}(\mu)$ according to Eq. (\ref{eq3.16}). We shall defer
the accomplishment of this task to the next Section, leaving the
matter as it is at this point, and proceed by considering the
remaining differential relation for the angular dependences of the
even multipole solutions we derived at the beginning of the present
Section.

\vv

\nd \textbf{(3) The inhomogeneous equation for
$f_\ell^{(n)}(\mu)$, $\ell = n+1, n+2, \ldots, 2n$}

For the treatment of Eq. (\ref{eq3.3}) it is useful to introduce the
differential operator: \beq \calt_\ell \equiv
(1-\mu^2)\frac{\dif^2}{\dif\mu^2} - \mu\frac{\dif}{\dif\mu} +
\ell^2\ , \label{e3.19} \eeq which has the property: \beq
\calt_k\left\{T_m(\mu)\right\} = (k^2 - m^2)T_m(\mu) \ \ \ (m = 0,
1, 2, \ldots), \label{eq3.20} \eeq as it can be easily proved with
the help of Eq. (\ref{eq3.11}). Recalling the definition of the
operator $\calh_\ell$ from Eq. (\ref{eq3.10}), we may restate Eq.
(\ref{eq3.3}) as: \beq \calt_\ell\left\{f_\ell^{(n)}(\mu)\right\} =
-\calh_{\ell-1}\left\{f_{\ell-1}^{(n)}(\mu)\right\} \ \ \ (\ell =
n+1, n+2, \ldots, 2n-1, 2n). \label{eq3.21} \eeq

The first use we shall give to Eq. (\ref{eq3.21}) is to find the
general form the angular functions $f_\ell^{(n)}(\mu)$ must obey.
For this purpose we shall view it as an inhomogeneous equation for
the function $f_\ell^{(n)}(\mu)$, therefore assuming that the
right hand side is known. We shall proceed by induction, starting
by considering the equation for the index $\ell = n+1$, which is:
\bey \calt_{n+1}\left\{f_{n+1}^{(n)}(\mu)\right\} &=&
-\calh_n\left\{f_n^{(n)}(\mu)\right\} \nonumber \\
&=& nA_{n,n}T_{n-1}(\mu)\ , \label{eq3.22} \eey the second
equality coming from the property of the operator $\calh_n$ stated
in Eq. (\ref{eq3.15}) when the object function is $f_n^{(n)}(\mu)$
as given by Eq. (\ref{eq3.4}).

Equation (\ref{eq3.22}) is an inhomogeneous Chebyshev differential
equation, which admits as complementary solution of even parity in
$\ta$ the Chebyshev polynomial type I of order $n+1$, and whose
particular solution should exhibit the same dependence on $\ta$ as
the driving term, being therefore proportional to the Chebyshev
polynomial of order $n-1$. We write then for the complete
solution: \beq f_{n+1}^{(n)}(\mu) = A_{n+1, n+1}^{(n)}T_{n+1}(\mu)
+ A_{n+1, n-1}^{(n)}T_{n-1}(\mu)\ , \label{eq3.23} \eeq where
$A_{n+1,n+1}^{(n)}$ and $A_{n+1,n-1}^{(n)}$ are constants.

We next consider Eq. (\ref{eq3.21}) with the index $\ell$ taken to
be equal to $n+2$, namely: \bey
&&\calt_{n+2}\left\{f_{n+2}^{(n)}(\mu)\right\} =
-\calh_{n+1}\left\{f_{n+1}^{(n)}(\mu)\right\} \nonumber \\
&&= \left[(n+1)A_{n+1,n+1}^{(n)} \!-\!
(2n-1)A_{n+1,n-1}^{(n)}\right]T_n(\mu)\! -\!
nA_{n+1,n-1}^{(n)}T_{n-2}(\mu), \label{eq3.24} \eey where we have
used the form just derived for $f_{n+1}^{(n)}(\mu)$ in the right
hand side and again the property of Eq. (\ref{eq3.15}) for the
differential operator $\calh_{n+1}$. The even component of the
complementary solution to Eq. (\ref{eq3.24}) is a multiple of the
Chebyshev polynomial type I of order $n+2$, while the particular
solution is a combination of the two Chebyshev polynomials that
appear on the right hand side. The complete solution of even parity
in the angle $\ta$ then writes as: \beq f_{n+2}^{(n)}(\mu) =
A_{n+2,n+2}^{(n)}T_{n+2}(\mu) + A_{n+2,n}^{(n)}T_n(\mu) +
A_{n+2,n-2}^{(n)}T_{n-2}(\mu)\ , \label{eq3.25} \eeq where the $A$'s
are constants.

From a consideration of the form of $f_n^{(n)}(\mu)$, given by Eq.
(\ref{eq3.4}), and of those of $f_{n+1}^{(n)}(\mu)$ and
$f_{n+2}^{(n)}(\mu)$, given respectively by Eqs. (\ref{eq3.23}) and
(\ref{eq3.25}), we infer that the form of the function
$f_{n+k}^{(n)}(\mu)$ in general must be: \beq f_{n+k}^{(n)}(\mu) =
\sum_{p=0}^kA_{n+k,n+k-2p}^{(n)}T_{n+k-2p}(\mu) \ \ \ (k = 0, 1, 2,
\ldots, n), \label{eq3.26} \eeq where the $A_{n+k,n+k-2p}^{(n)}$'s
are constants. Given that, according to Eq. (\ref{eq2.31}), the even
multipole solution of order $n$, $\vf^{(n)}(x,\mu)$, depends on
$n+1$ angular functions $f_i^{(n)}(\mu)$ ($i=n, n+1, \ldots, 2n$)
and since by Eq. (\ref{eq3.26}) each of these comprises $i-n+1$
constants $A_{i,j}^{(n)}$ ($j=i,i+1,\ldots, 2n-i$), the total number
of constants needed to specify $\vf^{(n)}(x,\mu)$ completely equals
\ $(n+1)(n+2)/2$, one of them remaining arbitrary, as it should be
proper to the solution of a homogeneous equation of the second order
with a definite parity; we shall take this constant, in terms of
which all others will be expressed, as $A_{2n,0}^{(n)}$. The problem
of finding the multipole solution of order $n$ to the sourceless
Grad-Shafranov equation is then the problem of determining the set
of these constants.

To deduce the recursion formulae connecting the coefficients that
enter the representation of $f_{n+k}^{(n)}$, we start by rewriting
Eq. (\ref{eq3.21}) as:
$$\calt_{n+k}\left\{f_{n+k}^{(n)}(\mu)\right\} =
-\calh_{n+k-1}\left\{f_{n+k-1}^{(n)}(\mu)\right\} \ \ \ (k = 0,
1,2, \ldots, n). \eqno(3.21')
$$

By inserting Eq. (\ref{eq3.26}) in the left hand side of this
equation and by using the property of Eq. (\ref{eq3.20}) for the
differential operator $\calt_{n+k}$, we obtain after a brief
calculation: \bey \calt_{n+k}\left\{f_{n+k}^{(n)}(\mu)\right\} &=&
4knA_{n+k,n-k}^{(n)}T_{n-k}(\mu) \!+\!
4\sum_{p=1}^{k-1}p(n\!+\!k\!-p)
A_{n+k,n+k-2p}^{(n)}T_{n+k-2p}(\mu) \nonumber \\
&&(k = 1, 2, \ldots, n). \label{eq3.27} \eey

The reduction of the right hand side of Eq. $(3.21')$ requires a
somewhat lengthier manipulation than that of the left hand side, but
it is otherwise straightforward. With the help of the formula stated
in Eq. (\ref{eq3.15}) we obtain: \bey
&&\hspace{-7mm}-\calh_{n+k-1}\left\{f_{n+k-1}^{(n)}(\mu)\right\} =
-(2k-3)nA_{n+k-1,n-k+1}^{(n)}T_{n-k}(\mu) \nonumber \\
&&\hspace{-7mm}-\sum_{p=1}^{k-1}\left[p(2n\!+\!2k\!-\!2p\!-\!3)A_{n\!+\!k\!-\!1,
n\!+\!k\!-\!2p\!-\!1}^{(n)} \!+\!(2p\!-\!3)(n\!+\!k\!-\!p)
A_{n\!+\!k\!-\!1,n\!+\!k\!-\!2p\!+\!1}^{(n)}\right]T_{n\!+\!k\!-\!2p}(\mu)
\nonumber \\
&&\hspace{-7mm}(k=1,2,\ldots, n). \label{eq3.28} \eey

The process of equating the left and the right hand sides of Eq.
$(3.21')$, as given by Eqs. (\ref{eq3.27}) and (\ref{eq3.28})
respectively, yields two recursion formulae for the coefficients
$A_{i,j}^{(n)}$. The first one comes from the terms proportional to
$T_{n-k}(\mu)$ on both sides of the equation and is: \beq A_{n+k-1,
n-k+1}^{(n)} = \left(\frac{4k}{3-2k}\right)A_{n+k,n-k}^{(n)}\ \ \
(k=n,n-1,n-2,\ldots, 2,1). \label{eq3.29} \eeq (The order in which
the values of $k$ are written above corresponds to the sequence in
which the values of the coefficients are generated by systematic
application of the formula, knowing the initial value
$A_{2n,0}^{(n)}$.)

The second recursion formula stems from the equality of the
coefficients multiplying $T_{n+k-2p}(\mu)$ on both sides of Eq.
$(3.21')$, and is: \bey &&A_{n+k-1,n+k-2p+1}^{(n)} =
-\frac{p}{2p\!-\!3}\left[4A_{n+k,n+k-2p}^{(n)} \!+\!
\left(\frac{2n\!+\!2k\!-\!2p\!-\!3}{n+k-p}\right)A_{n+k-1,n+k-2p-1}^{(n)}
\right] \nonumber \\
&&(k=n, n-1, \ldots, 3, 2; \ p = k-1, k-2, \ldots, 2,1).
\label{eq3.30} \eey

The set of difference equations constituted by Eq. (\ref{eq3.17})
together with Eq. (\ref{eq3.18}), Eq. (\ref{eq3.29}) and Eq.
(\ref{eq3.30}) encompasses all the coefficients that are needed to
specify the whole of the angular functions that enter the
composition of a multipole solution of a given order and forms thus
a complete formulation of the problem of finding the totality of
even polynomial solutions to Eq. (\ref{eq2.1}). In Section IV we
shall show how the difference equations can be used to generate the
coefficients by a step-by-step calculation procedure; next, partly
in Section IV and partly in Section V, we shall derive a solution in
closed form for each of the three difference equations, such that a
coefficient knowingly belonging to the scope of one of them can be
evaluated by the appropriate solving formula from the sole knowledge
of its indices; finally, also in Section V, we shall show that the
solutions found can be merged into a single formula, from which the
whole of the coefficients pertaining to a multipole solution can be
obtained in terms of $A_{2n,0}^{(n)}$, being enough for that to
specify the order of the multipole.

\vv\vv

\begin{center}
{\bf IV. THE TRIANGLE OF COEFFICIENTS AND THE SOLUTIONS FOR THE
PERIPHERAL COEFFICIENTS}
\end{center}

\setcounter{section}{4} \setcounter{equation}{0}

\vv

An insight into the mathematical structure of the problem posed by
the difference equations to which the original problem of the
partial differential equation for the multipole fields was reduced
in the last Section can be gained by displaying the Chebyshev
coefficients $A_{i,j}^{(n)}$ for the solution $\vf^{(n)}(x,\mu)$ of
a given order $n$ in a Cartesian array of columns and rows in which
the number attached to a column specifies the first suffix $i$ of
the coefficients keeping position in that column and the number
attached to a row specifies the second suffix $j$ of the
coefficients belonging to that row. According to Eq. (\ref{eq3.26})
the first suffix of the numerical coefficients $A_{i,j}^{(n)}$
belonging to the set of those characterized by possessing a
specified $i$ as the common first suffix is defined to be the
subscript of the angular function $f_i^{(n)}(\mu)$ into whose
composition they enter as factors multiplying Chebyshev polynomials,
and, by Eq. (\ref{eq2.31}), the subscripts of the angular functions
that are summoned to participate in the combination that builds up
the multipole solution $\vf^{(n)}(x,\mu)$ rank from $n$ to $2n$.
Thus the label of the columns in the array must run from $i=n$ to $i
= 2n$. With regard to the second suffix of the coefficient
$A_{i,j}^{(n)}$, which is equal to the order of the Chebyshev
polynomial this coefficient multiplies in Eq. (\ref{eq3.26}), it
must be clear from this equation that its range of variation,
considered in its wholeness the constellation of coefficients
involved in the construction of the multipole solution
$\vf^{(n)}(x,\mu)$, is determined by the angular function
$f_{2n}^{(n)}(\mu)$, the general expression of which is given by Eq.
(\ref{eq3.16}). From this we see that the label of the rows in the
array must run from $j=0$ to $j = 2n$.

Figure 2 exhibits such an array, in which $n$ is equal to 4 and the
coefficient $A_{8,0}^{(n)}$ is taken to be unity. As this example
illustrates, not all of the positions are to be filled, since, by
virtue of the conventions we have adopted regarding notation, there
are no coefficients associated with every pair of indices within the
wideness of range of the matrix $(i,j)$. In general, the field of
indices of the coefficients $A_{i,j}^{(n)}$ for a fixed $n$ is
defined by: \beq \left.\begin{array}{l} i = 2n,
2n-1, 2n-2, \ldots, n+1, n; \\
j = 2n-i, 2n-i+2, \ldots, i-2, i. \end{array}\right\} \label{eq4.1}
\eeq The area enclosing the coefficients resembles that of a
triangle, and for this reason we shall refer to this array as the
triangle of coefficients. In due time we shall show that it bears a
close kinship with the array of binomial coefficients known as
Pascal's triangle.

Each of the difference equations we have derived in Section III
applies to the coefficients occupying a different region of the
triangle according to a pattern of correspondence which we pass to
expound.

{\unitlength=1mm

\begin{center}

{\renewcommand {\arraystretch}{1.5} {\tabcolsep=1.5mm

\vspace{4cm}

\hspace{2.5cm}\begin{picture}(120,105)
\put(8,100){\begin{tabular}{|c||c|c|c|c|c|} \hline
\raisebox{-2mm}{$j$} \ \ \
\raisebox{2mm}{$i$} & 4 & 5 & 6 & 7 & 8  \\ \hline\hline 0 &&&&& 1  \\
\hline 1 &&&& $-\frac{16}{5}$ &  \\ \hline 2 & & & $\frac{64}{5}$ &
& $-\frac{56}{25}$  \\ \hline 3 & & $-\frac{512}{5}$ & &
$\frac{336}{25}$ &  \\ \hline 4 & $-\frac{2048}{5}\ $ & &
$-\frac{3584}{25}$ & & $\frac{84}{25}$
\\ \hline 5 & & $-\frac{3584}{5}\!\!$ & & $-\frac{336}{5}$ &  \\
\hline 6 & & & $\!\!-\frac{12096}{25}\!\!$ & & $-\frac{264}{25}$  \\
\hline 7 & & & & $-\frac{3696}{25}$ &  \\ \hline 8 & & & & &
$-\frac{429}{25}$   \\ \hline
\end{tabular}}
\put(88,128){\vector(0,-1){63}} \put(65,128){\vector(-3,-2){40}}
\put(19.7,132.5){\line(-3,2){11.5}} \put(71,117.5){\vector(3,-2){5}}
\put(75,109.5){\vector(-3,-2){5}}
\end{picture}

}}

\vspace{-5cm}

\begin{tabular}{lp{12cm}}
\textbf{FIG. 2} & The triangle of the coefficients $A_{i,j}^{(n)}$
for $n = 4$. The coefficient $A_{8,0}^{(4)}$ is taken to be unity.
The long arrows on the sides indicate the sequence in which the
peripheral coefficients are generated by recursion starting with the
coefficient $A_{8,0}^{(4)}$. The short interior arrows intend to
signify that the element $A_{7,1}^{(4)}$ combines with the element
$A_{8,2}^{(4)}$ to generate the coefficient $A_{7,3}^{(4)}$.
\end{tabular}

\end{center}

}

\vv

\vv

\nd\textbf{(1) The equation for the coefficients in the column on
the right hand side of the triangle of coefficients}

\vv

This is Eq. (\ref{eq3.18}) together with Eq. (\ref{eq3.17}), here
reproduced as: \vspace{2mm} \beq A_{2n,2\ell+2}^{(n)} =
-(1+\dt_{\ell, 0})\frac{(n-\ell)(2n+2\ell - 1)}{(n+\ell +
1)(2n-2\ell - 3)}A_{2n,2\ell}^{(n)} \ \ (\ell = 0, 1, 2, \ldots,
n-1), \label{eq4.1new} \eeq \vspace{2mm} \nd where we have made
use of the Kronecker symbol $\dt_{\ell, 0}$ in order to unify the
expressions for $\ell = 0$ and for $\ell \ne 0$. We find it useful
to define a new discrete variable $j$ relating to $\ell$ through:
\beq j = 2\ell + 2, \label{eq4.2} \eeq in terms of which Eq.
(\ref{eq4.1new}) is restated as: \vspace{2mm}\beq A_{2n,j}^{(n)}
\!= -(1+\dt_{j,2})\frac{(2n - j + 2)(2n + j - 3)}{(2n+j)(2n - j -
1)}A_{2n,j-2}^{(n)}\ \ (j \!=\! 2, 4, \ldots, 2n-2, 2n).
\label{eq4.3} \eeq

\vspace{2mm}

Note that, with this transformation, the free variable that appears
in the recursion formula for the coefficients is now denoted by the
same symbol that, as a suffix, indicates their positions in the
column $i = 2n$. By choosing any value for $A_{2n,0}^{(n)}$ (unity,
for example, as in the case illustrated by Fig. 2), all the
coefficients belonging to the column on the right hand side of the
triangle can be evaluated by setting $j = 2,4, \ldots$, up to $2n$
in succession. The arrow on the side of the column $i = 8$ in Fig. 2
indicates the sequence in which the positions are filled according
to this procedure.

A closed form solution to the homogeneous difference equation for
$A_{2n,j}^{(n)}$ can be easily obtained. Writing down Eq.
(\ref{eq4.3}) for $j = 2, 4, \ldots$ up to a generic (even) value
$j$ and then multiplying all the relations so obtained one by the
other in succession, we arrive at: \bey A_{2n,j}^{(n)} &=&
(-1)^{\frac{j}{2}} 2\frac{(2n)(2n-2) \cdots (2n-j+4)(2n-j+2)}{(2n+
2)(2n+4) \cdots (2n+j-2)(2n+j)} \nonumber \\
&&\times \frac{(2n-1)(2n+1) \cdots (2n+j-5)(2n+j-3)}{(2n-3)(2n-5)
\cdots (2n-j+1)(2n-j-1)} A_{2n,0}^{(n)} \nonumber \\
&&(j = 2,4, \ldots,2n). \label{eq4.4} \eey

This expression can be written in a more concise manner as: \bey
A_{2n,j}^{(n)} &=& (-1)^{\frac{j}{2}}\left[\frac{2n-1}{{2n\choose
n}}\right]^2\frac{2}{(2n-1-j)(2n-1+j)} {2n+j \choose
n+\frac{j}{2}} {2n-j \choose n-\frac{j}{2}} A_{2n,0}^{(n)}
\nonumber \\
&&(j = 2, 4, \ldots, 2n-2, 2n), \label{eq4.5} \eey where the symbol
${p\choose q}$ stands for the binomial coefficient of $p$ with
respect to $q$ as usually defined \cite{nove}. The constant
$A_{2n,0}^{(n)}$ remains arbitrary.

\vv

\nd \textbf{(2) The equation for the coefficients on the upper
side of the triangle of coefficients}

\vv

We now turn our attention to Eq. (\ref{eq3.29}) in Section III.
With the transformation of the independent variable: \beq k = i -
n + 1 \ , \label{eq4.6} \eeq this equation takes on the convenient
form: \beq A_{i,2n-i}^{(n)} = -4\left(\frac{i-n+1}{2i-2n-1}\right)
A_{i+1,2n-i-1}^{(n)} \ \ \ (i = 2n-1, 2n-2, \ldots,n),
\label{eq4.7} \eeq which can be recognized, the same as the
previous equation for the coefficients on the right hand side of
the triangle, as a first order, ordinary, homogeneous difference
equation for $A_{2n-j,j}^{(n)}$, whose solution depends on one
arbitrary constant. It is seen that, as the independent variable
is varied in accordance with the sequence of integers from $i =
2n-1$ to $i = n$, Eq. (\ref{eq4.7}) provides us with a recursive
scheme starting with $A_{2n,0}^{(n)}$ to evaluate the coefficients
whose positions are aligned along the top side of the triangle. In
Fig. 2 the sequence in which the coefficients for the case $n = 4$
are generated is indicated by an arrow above the upper side of the
triangle, the initiating coefficient being $A_{8,0}^{(4)} = 1$.

A closed form solution for Eq. (\ref{eq4.7}) can be derived by the
usual procedure of writing down the equations for $i = 2n-1, 2n-2,
\ldots$, down to a generic $i$ and then multiplying all of them
together. The result is: \bey A_{i,2n-i}^{(n)} &=& (-4)^{2n-i}
\frac{n(n-1)(n-2) \cdots (i-n+2)(i-n+1)}{(2n-3)(2n-5) \cdots
(2i-2n+1)(2i-2n-1)}A_{2n,0}^{(n)} \nonumber \\
&&(i =2n-1, 2n-2, \ldots, n). \label{eq4.8} \eey With the help of
the symbol for the binomial coefficients this expression can be
recast as \bey A_{i,2n-i}^{(n)} &=&
(-1)^i2^{3(2n-i)}\left(\frac{2n-1}{2i-2n-1}\right) \frac{{2i-2n
\choose i-n}}{{2n\choose n}} A_{2n,0}^{(n)} \nonumber \\
&&(i = 2n-1, 2n-2, \ldots, n+1, n). \label{eq4.9} \eey

\vv

\nd\textbf{(3) The equation for the internal coefficients}

\vv

We shall call internal coefficients all those that belong neither to
the column on the right hand side nor to the upper side of the
triangle of coefficients, including thus under this denomination
also the coefficients that fill in the positions along the down
side. For all of them the governing equation is Eq. (\ref{eq3.30}).
If we introduce the transformation defined by \beq \left.
\begin{array}{l} k = i-n+1\ , \\
\\
{\dps p =  1 + \meio(i-j)}\ , \end{array} \right\} \label{eq4.10}
\eeq it can be brought to the form: \bey
&&\hspace{-7mm}A_{i,j}^{(n)} = -\left(\frac{i-j+2}{i-j-1}\right)
\left[\left(\frac{i+j-3}{i+j}\right)A_{i,j-2}^{(n)} +
2A_{i+1,j-1}^{(n)}\right] \nonumber \\
&&\hspace{-7mm}(i \!=\! 2n\!-\!1, 2n\!-\!2, \ldots, n\!+\!2,
n\!+\!1; j \!=\! 2n\!-\!i\!+\!2, 2n\!-\!i\!+\!4, \ldots, i\!-\!2,
i). \label{eq4.12} \eey

The use of Eq. (\ref{eq4.12}) as a recursion formula to evaluate the
internal coefficients and complete the filling in of the positions
still vacant in the triangle requires that the positions in the
column on the right hand side and those along the upper skew side be
already filled up. For illustration of this requirement and of the
computational scheme brought forth by the above mentioned difference
equation, the arrows in Fig. 2 directed from the element
$A_{7,1}^{(4)}$ to the element $A_{8,2}^{(4)}$ and from this one to
the element $A_{7,3}^{(4)}$ intend to indicate that the last-named
follows up an operation performed on the two first.

Distinctly from the two other previously considered difference
equations, Eq. (\ref{eq4.12}) is a partial difference equation and
one of the second order, since it relates the unknown function
$A_{i,j}^{(n)}$ in three neighboring positions not aligned on the
plane $(i,j)$. Two of the positions, however, belong to the same
column and this makes it possible to treat the equation as an
ordinary difference equation of the first order in the variable
designating the row for the coefficients belonging to this common
column, the free variable designative of the column itself being
seen as a parameter. This approach requires that the third
coefficient present in the equation, whose place in the $(i,j)$
diagram falls on a neighboring column, be assumed to be known, and
leads to a recursion relation between columns rather than a relation
between single elements, as in Eq. (\ref{eq4.12}).

We rewrite Eq. (\ref{eq4.12}) under the form of an inhomogeneous
equation as:
$$A_{i,j}^{(n)} - h(i,j-2)A_{i,j-2}^{(n)} =
g(i+1,j-1)A_{i+1,j-1}^{(n)}, \eqno(4.12')
$$
where we have introduced the functions: \beq h(i,j) =
-\frac{(i-j)(i+j-1)}{(i-j-3)(i+j+2)} \label{eq4.13} \eeq and \beq
g(i,j) = -2\left(\frac{i-j}{i-j-3}\right)\ , \label{eq4.14} \eeq and
where the right hand side is taken as the forcing term, supposed to
be known. The resolution of this difference equation requires, as a
first step, that it be multiplied by a ``summing factor''
\cite{onze}, the analog of an integrating factor for a first order
differential equation, the effect of which is to convert the left
hand side into an exact ``discrete differential''. To find out which
is this factor we multiply Eq. $(4.12')$ by the reciprocal of a
function $W^{(n)}(i,j)$, unknown as yet, and obtain: \beq
\frac{A_{i,j}^{(n)}}{W^{(n)}(i,j)} -
\frac{h(i,j-2)A_{i,j-2}^{(n)}}{W^{(n)}(i,j)} =
\frac{g(i+1,j-1)A_{i+1, j-1}^{(n)}}{W^{(n)}(i,j)}\ . \label{eq4.15}
\eeq

To make the left hand side assume the desired form, we must choose
$W^{(n)}(i,j)$ in such way that the equality: \bey W^{(n)}(i,j) =
h(i,j-2)W^{(n)}(i,j-2) && \!\!\!(i = 2n-1, 2n-2, \ldots, n+1;
\nonumber
\\
&&\!\!\!j \!=\! 2n\!-\!i\!+\!2, 2n\!-\!i\!+\!4, \ldots, i)
\label{eq4.16} \eey be true, case in which Eq. (\ref{eq4.15}) can be
written as: \beq a_{i,j}^{(n)} - a_{i,j-2}^{(n)} = g(i+1,j-1)
\frac{W^{(n)}(i+1,j-1)}{W^{(n)}(i,j)} a_{i+1,j-1}^{(n)}\ ,
\label{eq4.17} \eeq where we have employed the notation: \beq
a_{i,j}^{(n)} \equiv \frac{A_{i,j}^{(n)}}{W^{(n)}(i,j)} \ .
\label{eq4.18} \eeq

The condition imposed upon $W^{(n)}(i,j)$, which translates by Eq.
(\ref{eq4.16}), can be viewed as a recursion formula for the
dependence of $W^{(n)}(i,j)$ on the variable $j$. No demand is made
on its end value, which can be chosen \textit{ad libitum}, and
regarding the convenience coming from simplicity, we take \beq
W^{(n)}(i,2n-i) = 1 \ . \label{eq4.19} \eeq

The solution to Eq. (\ref{eq4.16}) then flows from the method
generally applicable to first order, homogeneous difference
equations, and is: \beq W^{(n)}(i,j) = h(i,2n-i)h(i, 2n-i+2)
\cdots h(i,j-2)\ . \label{eq4.20} \eeq

By use of Eq. (\ref{eq4.13}) for $h(i,j)$, and making appeal to the
symbol of binomial coefficients to give a concise representation to
the product on the right hand side, this expression can be brought
to the form: \bey &&W^{(n)}(i,j) = (-1)^{\frac{i+j}{2}-n}
\frac{2n-1}{{2n\choose n}} \frac{2i-2n-1}{(i - j -1)(i+j-1)}
\frac{{i+j \choose \frac{i + j}{2}}{i-j \choose \frac{i-j}{2}}}{
{2i-2n\choose i-n}}
\nonumber \\
&&(i=2n-1, 2n-2, \ldots, n+1; j = 2n-i + 2, 2n-i+4, \ldots, i).
\label{eq4.21} \eey

With the help of this formula, the factor that appears on the right
hand side of Eq. (\ref{eq4.17}) can be readily shown to be: \beq
\frac{W^{(n)}(i+1,j-1)}{W^{(n)}(i,j)} = -\frac{4}{g(i+1,j-1)}
\left(\frac{i-n+1}{2i-2n-1}\right)\ , \label{eq4.22} \eeq where we
have made use of the notation for the function $g(i,j)$ introduced
by Eq. (\ref{eq4.14}).

We thus have for the weighted coefficients $a_{i,j}^{(n)}$ the
equation: \bey &&a_{i,j}^{(n)} - a_{i,j-2}^{(n)} = -4\left(\frac{i
- n + 1}{2i-2n-1}\right) a_{i+1,j-1}^{(n)} \nonumber \\
&&(i = 2n-1, 2n-2, \ldots, n+1; j = 2n-i+2, 2n-i+4, \ldots, i).
\label{eq4.23} \eey

Considered as an ordinary difference equation for the elements
aligned along the column labelled by $j$ in the triangle of
coefficients, the end condition for Eq. (\ref{eq4.23}) is provided
by Eq. (\ref{eq4.7}), here rewritten as:
$$a_{i,2n-i}^{(n)} = -4\left(\frac{i-n+1}{2i-2n-1}\right)a_{i+1,
2n-i-1}^{(n)}\ \ \ (i=2n-1, 2n-2, \ldots, n), \eqno (4.8')
$$
where we have employed the equality between the weighted
coefficients and the coefficients themselves at the positions on
the top of the columns: \beq a_{k,2n-k}^{(n)} = A_{k, 2n-k}^{(n)}\
, \label{eq4.24new} \eeq which is itself a consequence of the
adoption of Eq. (\ref{eq4.19}) as end condition for
$W^{(n)}(i,j)$.

By writing Eq. (\ref{eq4.23}) for a fixed $i$ and $j = 2n-i+2,
2n-i+4, \ldots$ up to a generic $j$ in succession, and then adding
all the equations so obtained, we arrive at the solution for the
weighted coefficients $a_{i,j}^{(n)}$ belonging to the $i$-column in
terms of a summation carried on coefficients belonging to the
neighboring $i+1$-column: \bey &&\hspace{-5mm}a_{i,j}^{(n)} =
-4\left(\frac{i-n+1}{2i-2n-1}\right)\sum_{\ell = 0}^{\frac{i+j}{2}
- n} a_{i+1, 2n-i-1+2\ell}^{(n)} \nonumber \\
&&\hspace{-5mm}(i = 2n-1,2n-2, \ldots, n+1, n; j = 2n\!-\!i,
2n\!-\!i\!+\!2, \ldots, i\!-\!2, i). \label{eq4.24} \eey

Regarding future use, it is also of interest to point out here
that, from Eqs. (\ref{eq4.18}), (\ref{eq4.5}) and (\ref{eq4.21}),
the weighted coefficients for the column on the right hand side of
the triangle take on the values: \setcounter{equation}{0}
\renewcommand{\theequation}{4.26\alph{equation}}
\bey a_{2n,0}^{(n)} &=& A_{2n,0}^{(n)} \
, \label{eq4.25a} \\
a_{2n,j}^{(n)} &=& 2A_{2n,0}^{(n)} \ \ (j = 2, 4, \ldots, 2n).
\label{eq4.25b} \eey

\setcounter{equation}{26}
\def\theequation{\thesection.\arabic{equation}}

We conclude this Section by illustrating the use of the several
recursion formulae for the coefficients we have derived by means of
a numerical application.

\vv

\nd \textit{\textbf{Example.} Evaluation of the Chebyshev
coefficients for the multipole solution of order \ $n = 4$.}

\vv

\nd (a) We start by the coefficients belonging to the column on
the right hand side of the triangle of coefficients. The recursion
formula to be employed in this case is that of Eq. (\ref{eq4.3}),
which, for $n = 4$, is written as: \beq A_{8,j}^{(4)} =
-(1+\dt_{j,2}) \frac{(10-j)(5+j)}{(8+j)(7-j)} A_{8,j-2}^{(4)} \ \
(j = 2, 4, 6, 8). \label{eq4.27} \eeq

Assuming that $A_{8,0}^{(4)} = 1$, we obtain in succession: \beq
A_{8,2}^{(4)} = -\frac{56}{25}, \ A_{8,4}^{(4)} = \frac{84}{25}, \
A_{8,6}^{(4)} = -\frac{264}{25}, \ A_{8,8}^{(4)} =
-\frac{429}{25}\ . \label{eq4.28} \eeq

\vv

\nd (b) We now evaluate the coefficients for the column adjacent
to that on the right hand side of the triangle, using Eq.
(\ref{eq4.23}) as a recurrence formula between contiguous rows.
For $n = 4$ and $i = 7$, this is: \beq a_{7,j}^{(4)} =
a_{7,j-2}^{(4)} - \frac{16}{5}a_{8,j-1}^{(4)}  \ \ (j = 3,5, 7).
\label{eq4.29} \eeq

The weighted coefficient on the top of the column must be
evaluated from Eq. (\ref{eq4.24}), which furnishes: \beq
a_{7,1}^{(4)} = -\frac{16}{5}a_{8,0}^{(4)}\ . \label{eq4.30} \eeq

The other ones that are required in precedence to the use of Eq.
(\ref{eq4.29}) are those making up the column corresponding to $i =
8$ and are promptly given by Eqs. (\ref{eq4.25a}) and
(\ref{eq4.25b}). We have: \beq a_{8,0}^{(4)} = 1, \ a_{8,2}^{(4)} =
a_{8,4}^{(4)}= a_{8,6}^{(4)} = a_{8,8}^{(4)} = 2\ . \label{eq4.31}
\eeq

We then obtain from Eqs. (\ref{eq4.30}), (\ref{eq4.31}) and from
the repeated use of Eq. (\ref{eq4.29}): \beq a_{7,1}^{(4)} =
-\frac{16}{5}, \ a_{7,3}^{(4)} = -\frac{48}{5}, \ a_{7,5}^{(4)} =
-16, \ a_{7,7}^{(4)} = -\frac{112}{5}\ . \label{eq4.32} \eeq

The bridge between the weighted coefficients $a_{7,j}^{(4)}$ and the
coefficients $A_{7,j}^{(4)}$ is the function $W^{(4)}(7,j)$, the
knowledge of which requires the knowledge of the function $h(7,j)$.
By resorting to Eq. (\ref{eq4.13}) we obtain for the values of $j$
of interest: \beq h(7,1) = -\frac{7}{5}\ , h(7,3) = -3\ , h(7,5) =
\frac{11}{7}\ . \label{eq4.33} \eeq

With this, the values of $W^{(4)}(7,j)$ can be computed
recursively from Eq. (\ref{eq4.16}), which, for $i = 7$, becomes:
\beq W^{(4)}(7,j) = h(7,j-2)W^{(4)}(7,j-2)\ . \label{eq4.34} \eeq
Starting with: \beq W^{(4)}(7,1) = 1\ , \label{eq4.35} \eeq in
obedience to the end condition stated in Eq. (\ref{eq4.19}), now
obligatory as a matter of consistency with the values ascribed to
$a_{8,j}^{(4)}$ in Eq. (\ref{eq4.31}), by putting $j = 3,5$ and 7
in Eq. (\ref{eq4.34}) we obtain one after the other: \beq
W^{(4)}(7,3) = -\frac{7}{5}, \  W^{(4)}(7,5) = \frac{21}{5}, \
W^{(4)}(7,7) = \frac{33}{5}\ . \label{eq4.36} \eeq

The values of the Chebyshev coefficients $A_{7,j}^{(4)}$,
according to Eq. (\ref{eq4.18}), are given by the product of the
weighted coefficients $a_{7,j}^{(4)}$ and the values of the
function $W^{(4)}(7,j)$, and this leads to: \beq A_{7,1}^{(4)} =
-\frac{16}{5}, \ A_{7,3}^{(4)} = \frac{336}{25}, \ A_{7,5}^{(4)} =
-\frac{336}{5}, \ A_{7,7}^{(4)} = -\frac{3696}{25}\ .
\label{eq4.37} \eeq

\vv

\nd (c) We now complete the filling in of the vacancies in the
triangle of coefficients using, for the purpose of illustration, the
two other recursion formulae we have derived, namely, the one given
by Eq. (\ref{eq4.7}), which applies to the coefficients high up on
the columns, and that of Eq. $(4.12')$, which generates the internal
coefficients. For $n = 4$ the first of these is: \beq
A_{i,8-i}^{(4)} = -4\left(\frac{i-3}{2i-9}\right)A_{i+1, 7-i}^{(4)}
\ \ (i=7, 6, 5, 4), \label{eq4.38} \eeq and gives, in addition to
the value of $A_{7,1}^{(4)}$ already found, the following ones: \beq
A_{6,2}^{(4)} = \frac{64}{5}, \ A_{5,3}^{(4)} = -\frac{512}{5}, \
A_{4,4}^{(4)} = -\frac{2048}{5}\ . \label{eq4.39} \eeq

For the column $i = 6$, Eq. $(4.12')$ becomes for $j = 4$ and $j =
6$, respectively: \bey A_{6,4}^{(4)} &=& h(6,2)A_{6,2}^{(4)}
+ g(7,3)A_{7,3}^{(4)} \ , \label{eq4.40} \\
A_{6,6}^{(4)} &=& h(6,4)A_{6,4}^{(4)} + g(7,5)A_{7,5}^{(4)}\ .
\label{eq4.41} \eey

Referring to Eqs. (\ref{eq4.13}) and (\ref{eq4.14}) we determine:
\beq \left. \begin{array}{l} {\dps h(6,2) =
-\frac{14}{5}, \ h(6,4) = \frac{3}{2}} \ ; \\
\\
{\dps g(7,3) = -8, \ g(7,5) = 4}\ , \end{array}\right\}
\label{eq4.42} \eeq from which and from the values already known
for the coefficients on the right hand side of Eqs. (\ref{eq4.40})
and (\ref{eq4.41}) we obtain: \beq A_{6,4}^{(4)} =
-\frac{3584}{25}, \ A_{6,6}^{(4)} = -\frac{12096}{25}\ .
\label{eq4.43} \eeq

Finally, putting $i = 5$, $j = 5$ in Eq. $(4.12')$ we have the
relation for the last coefficient still missing in the triangle:
\beq A_{5,5}^{(4)} = h(5,3)A_{5,3}^{(4)} + g(6,4)A_{6,4}^{(4)}\ ,
\label{eq4.44} \eeq which, with \beq h(5,3) = \frac{7}{5} \ \ \hbox{
and } \ \ g(6,4) = 4\ , \label{eq4.45} \eeq yields: \beq
A_{5,5}^{(4)} = -\frac{3584}{5}\ . \label{eq4.46} \eeq

Having found all the Chebyshev coefficients, we can now establish
the multipole solution of the fourth order, whose form, according
to Eq. (\ref{eq2.31}), is \beq \vf^{(4)}(x,\mu) =
f_4^{(4)}(\mu)x^4 + f_5^{(4)}(\mu)x^5 + f_6^{(4)}(\mu)x^6 +
f_7^{(4)}(\mu)x^7 + f_8^{(4)}(\mu)x^8\ . \label{eq4.47} \eeq

Considering that the Chebyshev polynomial $T_n(\mu = \cos\ta)$ ($n =
0, 1, 2, \ldots$) is identical with the trigonometric function $\cos
n\ta$, and recalling the general expression for $f_{n+k}^{(n)}(\mu)$
as given by Eq. (\ref{eq3.26}), the angular functions that enter the
constitution of $\vf^{(4)}(x,\mu)$ of the present case can be
written as: \setcounter{equation}{0}
\renewcommand{\theequation}{4.48\alph{equation}}
\bey f_4^{(4)}(\mu) &=& A_{4,4}^{(4)}T_4(\mu) \nonumber \\
&=& -\frac{2048}{5}\cos 4\ta\ , \label{eq4.48a} \\
f_5^{(4)}(\mu) &=& A_{5,3}^{(4)}T_3(\mu) + A_{5,5}^{(4)}T_5(\mu)
\nonumber \\
&=& -\frac{512}{5}\cos 3\ta - \frac{3584}{5}\cos 5\ta\ ,
\label{eq4.48b} \\
f_6^{(4)}(\mu) &=& A_{6,2}^{(4)}T_2(\mu) + A_{6,4}^{(4)}T_4(\mu) +
A_{6,6}^{(4)}T_6(\mu) \nonumber \\
&=& \frac{64}{5}\cos 2\ta - \frac{3584}{25}\cos 4\ta -
\frac{12096}{25}\cos 6\ta\ , \label{eq4.48c} \\
f_7^{(4)}(\mu) &=& A_{7,1}^{(4)}T_1(\mu) + A_{7,3}^{(4)}T_3(\mu) +
A_{7,5}^{(4)}T_5(\mu) + A_{7,7}^{(4)}T_7(\mu) \nonumber \\
&=& -\frac{16}{5}\cos\ta + \frac{336}{25}\cos 3\ta - \frac{336}{5}
\cos 5\ta - \frac{3696}{25}\cos 7\ta\ , \label{eq4.48d} \\
f_8^{(4)}(\mu) &=& A_{8,0}^{(4)}T_0(\mu) + A_{8,2}^{(4)}T_2(\mu) +
A_{8,4}^{(4)}T_4(\mu) + A_{8,6}^{(4)}T_6(\mu) +
A_{8,8}^{(4)}T_8(\mu) \nonumber \\
&=& 1 - \frac{56}{25}\cos 2\ta + \frac{84}{25}\cos 4\ta -
\frac{264}{25}\cos 6\ta - \frac{429}{25}\cos 8\ta\ . \label{eq4.48e}
\eey

Alternatively, using \cite{nove}, \cite{doze}:
\setcounter{equation}{48}
\def\theequation{\thesection.\arabic{equation}}
\bey T_3(\mu) &=& -3\mu + 4\mu^3\ , \label{eqnova4.49} \\
T_4(\mu) &=& 1 - 8\mu^2 + 8\mu^4\ ,
\label{eq4.48} \\
T_5(\mu) &=& 5\mu - 20\mu^3 +\mu^5\ , \label{eq4.49} \\
T_6(\mu) &=& -1 + 18\mu^2 - 48\mu^4 + 32\mu^6 \ , \label{eq4.50}
\\
T_7(\mu) &=& -7\mu + 56\mu^3- 112\mu^5 + 64\mu^7\ , \label{eq4.51}
\\
T_8(\mu) &=& 1 - 32\mu^2 + 160\mu^4 - 256\mu^6 + 128\mu^8\ ,
\label{eq4.52} \eey and $T_1(\mu)$ and $T_2(\mu)$ as given by Eqs.
(\ref{eq2.9}) and (\ref{eq2.21}), these same angular functions can
be stated in the form: \setcounter{equation}{0}
\renewcommand{\theequation}{4.55\alph{equation}}
\bey f_4^{(4)}(\mu) &=& \frac{2048}{5}(-1+8\mu^2 - 8\mu^4)\ , \label{eq4.53a} \\
f_5^{(4)}(\mu) &=& \frac{4096}{5}(-4\mu + 17\mu^3 - 14\mu^5)\ ,
\label{eq4.53b} \\
f_6^{(4)}(\mu) &=& \frac{1024}{25}(8-184\mu^2 + 539\mu^4 -
378\mu^6)\ , \label{e4.53c} \\
f_7^{(4)}(\mu) &=& \frac{1024}{25}(16\mu - 168\mu^3 + 378\mu^5 -
231\mu^7)\ , \label{eq4.53d} \\
f_8^{(4)}(\mu) &=& \frac{128}{25}(64\mu^2 - 432\mu^4 + 792\mu^6 -
429\mu^8)\ . \label{eq4.53e} \eey

\setcounter{equation}{55}
\def\theequation{\thesection.\arabic{equation}}

Combining the angular functions $f_4^{(4)}(\mu), f_5^{(4)}(\mu),
\ldots, f_8^{(4)}(\mu)$ listed above according to Eq.
(\ref{eq4.47}), the multipole solution of order $n = 4$ is obtained.
In Appendix A and in Appendix B the expressions for $\vf^{(4)}(x,
\ta)$ and for $\vf^{(4)}(x,\mu)$, as combinations of harmonics of
the angle $\ta$ and as polynomials in $\mu$ respectively, are
written in full after being divided by a normalization factor equal
to 512.

\vv

\begin{center}
{\bf V. SOLUTION TO THE DIFFERENCE EQUATION FOR THE \\
INTERNAL COEFFICIENTS. UNIFIED EX\-PRES\-SION \\
FOR THE CHEBYSHEV COEFFICIENTS OF THE \\
MULTIPOLE SOLUTIONS}
\end{center}

\setcounter{section}{5} \setcounter{equation}{0}

\vv

In the previous Section the problem of the determination of the
Chebyshev coefficients $A_{i,j}^{(n)}$ for the multipole solution of
order $n$ to the sourceless Grad-Shafranov equation was reduced to
that of a partial difference equation of the second order subjected
to boundary conditions on two frontiers of the domain of the two
variables $i$ and $j$ in which the problem is stated, one
corresponding to the ``line'' $i+j = 2n$ and the other to the
``line'' $i = 2n$. The boundary conditions take the form of values
imposed on the dependent variable which are themselves fixed by two
ordinary difference equations, for which solutions were obtained in
terms of a single coefficient which remains arbitrary. In the
present Section we establish the solution to the partial difference
equation apt to the inner region and show that it makes possible to
express the coefficients $A_{i,j}^{(n)}$ by a single formula which
encompasses all domains pertaining to the statement of the problem,
the border sites and the inside alike.

We commence by considering Eq. (\ref{eq4.23}) for the weighted
coefficients $a_{i,j}^{(n)}$, in which only one of the multiplying
coefficients to the unknown function in the three terms of which it
is constituted is not a constant. The fact that this coefficient
depends on only one of the two independent variables opens the way
to converting the equation into one of constant coefficients.
Indeed, if we write the dependent variable as: \beq a_{i,j}^{(n)} =
p^{(n)}(i)b_{i,j}^{(n)}\ , \label{eq5.1} \eeq Eq. (\ref{eq4.23})
becomes: \bey &&p^{(n)}(i)\left[b_{i,j}^{(n)} -
b_{i,j-2}^{(n)}\right] =
-4\left(\frac{i-n+1}{2i-2n-1}\right)p^{(n)}(i+1)b_{i+1,j-1}^{(n)}
\nonumber \\
&&(i = 2n-1, 2n-2, \ldots, n+1; \ j = 2n-i +2, 2n-i+4, \ldots, i).
\label{eq5.2} \eey If now we put: \beq p^{(n)}(i) =
-4\left(\frac{i-n+1}{2i-2n-1}\right)p^{(n)}(i+1), \label{eq5.3} \eeq
then the equation governing the transformed function $b_{i,j}^{(n)}$
is: \bey &&b_{i,j}^{(n)} - b_{i,j-2}^{(n)} -
b_{i+1, j-1}^{(n)} = 0 \nonumber \\
&&(i=2n-1, 2n-2, \ldots, n+1; \ j = 2n-i+2, 2n-i+4, \ldots, i),
\label{eq5.4} \eey which bears the feature of being of constant
coefficients.

We first take care of Eq. (\ref{eq5.3}), which can be easily solved.
Writing it for $i = 2n-1, 2n-2, \ldots$, down to a generic $i \ge
n+1$, and then multiplying the equations so obtained one by the
other all together, we reach: \bey &&p^{(n)}(i) =
(-1)^{2n-i}2^{3(2n-i)}\left(\frac{2n-1}{2i-2n-1}\right)
\frac{{2i-2n\choose i-n}}{{2n\choose n}} p^{(n)}(2n) \nonumber \\
&&(i = 2n-1, 2n-2, \ldots, n+1). \label{eq5.5} \eey

We are free to choose the end value for $p^{(n)}(i)$ and in the
benefit of simplicity we put $p^{(n)}(2n) = 1$.

The attack to Eq. (\ref{eq5.4}) has to be preceded by the
specification of the boundary conditions to be applied to the
function $b_{i,j}^{(n)}$. For the column making the right hand side
frontier of the domain of the free variables, \textit{id est}, for
$i = 2n$, from Eq. (\ref{eq4.25b}) and Eq. (\ref{eq5.1}) we have:
\beq b_{2n,j}^{(n)} = 2A_{2n,0}^{(n)} \ \ (j = 2,4, \ldots, 2n).
\label{eq5.6} \eeq

For the positions high up in the columns, having recourse to the
connection between $a_{2n-j, j}^{(n)}$ and $b_{2n-j,j}^{(n)}$ coming
from Eq. (\ref{eq5.1}), then evaluating $p^{(n)}(2n-j)$ according to
Eq. (\ref{eq5.5}), and finally recalling the frontier values for the
weighted coefficients $a_{i,j}^{(n)}$ and for the coefficient
$A_{i,j}^{(n)}$ as they come stated in Eqs. (\ref{eq4.24new}) and
(\ref{eq4.9}) respectively, we are able to establish that: \beq
b_{2n-j,j}^{(n)} = A_{2n,0}^{(n)}\ \ \ (j = 0, 1, \ldots,
n).\label{eq5.7} \eeq

Having determined all the boundary conditions to be satisfied by the
solution of Eq. (\ref{eq5.4}) we turn our attention to the equation
itself. A more familiar form can be given to it by introducing new
independent variables  $k$ and $\ell$ related to the ones we have
been  using until now through the equations: \beq
\left. \begin{array}{l} {\dps k = n-\frac{i-j}{2}} \ ,  \\
{\dps \ell = \frac{i+j}{2} - n}\ .
\end{array}\right\} \label{eq5.8} \eeq

The ranges of variation of $k$ and $\ell$ are respectively: \beq
\left.\begin{array}{l}
k = 0, 1, 2, \ldots, n-1, n; \\
\ell = 0, 1, 2, \ldots, k-1,k \ . \end{array}\right\}
\label{eq5.9} \eeq

Displayed in a Cartesian arrangement, the domain of the pair of
variables $(k,\ell)$ still shows to be a triangular one but at
variance with that of the variables $(i,j)$, it contains no
positions associated with holes, since the allowed values for one
and the other variables obey a complete sequence of integers
(starting with zero). Note that the inverse transformation to the
one defined by Eq. (\ref{eq5.8}) is: \beq
\left. \begin{array}{l} i = 2n-k+\ell \ ,  \\
j = k + \ell\ . \end{array}\right\} \label{eq5.10} \eeq

Following the transformation of free variables $(i,j) \to (k,\ell)$,
the dependent variable transforms as $b_{i,j}^{(n)} \to
F_{k,\ell}^{(n)}$, and, in place of Eq. (\ref{eq5.4}) we have: \bey
&&F_{k,\ell}^{(n)} = F_{k-1,\ell-1}^{(n)} +
F_{k-1,\ell}^{(n)} \nonumber \\
&&(k = 2,3, \ldots, n-1, n; \ \ell = 1, 2, \ldots, k-1).
\label{eq5.11} \eey

After consultation to Eqs. (\ref{eq5.6}) and (\ref{eq5.7}) on the
boundary conditions for $b_{i,j}^{(n)}$, we find that they translate
for $F_{k,\ell}^{(n)}$ as: \setcounter{equation}{0}
\renewcommand{\theequation}{5.12\alph{equation}}
\beq F_{k,k}^{(n)} = 2A_{2n,0}^{(n)} \ \ \ (k=1, 2, \ldots,n)
\label{eq5.12} \eeq and \beq F_{k,0}^{(n)} = A_{2n,0}^{(n)} \ \ \
(k = 0,1, 2,\ldots, n). \label{eq5.13} \eeq
\setcounter{equation}{12}
\def\theequation{\thesection.\arabic{equation}}

The equation we have been able to derive for $F_{k,\ell}^{(n)}$
could be called Stifel's equation, since it bears the precise form
of the formula associated with the name of Stifel that relates three
contiguous binomial coefficients and that provides the basis for the
step-by-step procedure of construction of Tartaglia's triangle (also
known as Pascal's arithmetic triangle) \cite{dez} for the
coefficients of the binomial expansion. The solution of Eq.
(\ref{eq5.11}), however, is not a multiple of the binomial
coefficient ${k\choose \ell}$ because of the factor 2 multiplying
the arbitrary constant $A_{2n,0}^{(n)}$ that appears in Eq.
(\ref{eq5.12}) for the values of $F_{k,\ell}^{(n)}$ along the
boundary $\ell = k$. Perhaps the easiest way to establish the
solution to the problem we have in hand is by inspecting a version
of Pascal's  triangle that associates two different constant values
with the two lines that delimit the available area for its expansion
respectively, distinctly from the classical version in which just a
single value is assumed for both.

Consider a quantity $G_{k,\ell}$ ($k = 2, 3, \ldots; \ell = 1, 2,
\ldots, k-1$) that obeys Stifel's relation the same as
$F_{k,\ell}^{(n)}$ in Eq. (\ref{eq5.11}), and that takes on the
value $G_{k, \ell} = 1 + c$, $c$ being an arbitrary constant, at the
side boundary $\ell = k$ ($k \ge 1$) of the domain of the free
variables, while keeping the reference value $G_{k,\ell} = 1$ along
the upper boundary ($\ell = 0$, $k = 0, 1, 2,\ldots$). A limited
extension of the arithmetic triangle for this case is represented in
Fig. 3.

Each position in the table is filled by summing the constituents in
two neighboring positions, both of which in the column immediately
to the left side of that position, one in the upper row, and the
other, below the latter, in the same row. The value of the entry in
a position $(k,\ell)$ that is obtained by following this rule is
made up of two parcels which can be traced back each to two
analogous sets of starting values respectively, but imposed on
displaced boundaries. The first parcel is the same that we would
have with $c = 0$ and there is no need to say more about it than
that it is the binomial coefficient ${k \choose \ell}$ of the
classical Pascal's triangle. The second parcel, which contains the
constant $c$ as a factor, is obtained by filling all positions in
the row below the uppermost one having $k \ge 1$ with the constant
value $c$ and the ones along the side boundary having $\ell \ge 2$
also with $c$, and applying the step-by-step building procedure of
the table thereafter. This means that the number multiplying $c$ in
the position $(k, \ell)$ is still a binomial coefficient, but
shifted with respect to the one unrelated to $c$ by one column and
one row, namely, it is the coefficient ${k-1\choose \ell - 1}$. The
addition of the two parcels gives then the tabulated quantity in
Fig. 3 as: \beq G_{k,\ell} = {k\choose \ell} + c{k-1 \choose \ell -
1}, \label{eq5.14} \eeq or, which is the same, as: \beq G_{k,\ell} =
\left(\frac{k+c\ell}{k}\right) {k\choose \ell}\ . \label{eq5.15}
\eeq

{\unitlength=1mm

\begin{center}

{\renewcommand {\arraystretch}{1.5} {\tabcolsep=0.9mm

\hspace{3cm}\begin{picture}(120,110)
\put(-28,57){\begin{tabular}{|c||c|c|c|c|c|c|c|c|c|c|c|} \hline
\raisebox{-2mm}{$\ell$} \ \ \
\raisebox{2mm}{$k$} & 0 & 1 & 2 & 3 & 4 & 5 & 6 & 7 & 8 & 9 & \ $\cdots$ \ \\
\hline\hline 0 & \ \ \,\footnotesize{1} \ \  & \footnotesize{1} &
\footnotesize{1} & \footnotesize{1} & \footnotesize{1} &
\footnotesize{1} & \footnotesize{1} & \footnotesize{1} &
\footnotesize{1} & \footnotesize{1} & $\cdots$
\\ \hline
1 & & \footnotesize{$1+c$} & \footnotesize{$2+c$} &
\footnotesize{$3+c$} & \footnotesize{$4+c$} & \footnotesize{$5+c$}
& \footnotesize{$6+c$} & \footnotesize{$7+c$} &
\footnotesize{$8+c$} & \footnotesize{$9+c$} & $\cdots$
\\ \hline 2 & & & \footnotesize{$1+c$} & \footnotesize{$3+2c$} &
\footnotesize{$6+3c$} & \footnotesize{$10+4c$} & \footnotesize{$15+5c$} & \footnotesize{$21+6c$} & \footnotesize{$28+7c$} & \footnotesize{$36+8c$} & $\cdots$   \\
\hline 3 & & & & \footnotesize{$1+c$} & \footnotesize{$4+3c$} & \footnotesize{$10+6c$} & \footnotesize{$20+10c$} & \footnotesize{$35+15c$} & \footnotesize{$56+21c$} & \footnotesize{$84+28c$} & $\cdots$   \\
\hline 4 & & & & & \footnotesize{$1+c$} & \footnotesize{$5+4c$} & \footnotesize{$15+10c$} & \footnotesize{$35+20c$} & \footnotesize{$70+35c$} & \footnotesize{$126+56c$} & $\cdots$   \\
\hline 5 & & & & & & \footnotesize{$1+c$} & \footnotesize{$6+5c$} & \footnotesize{$21+15c$} & \footnotesize{$56+35c$} & \footnotesize{$126+70c$} & $\cdots$   \\
\hline
6 & & & & & & & \footnotesize{$1+c$} & \footnotesize{$7+6c$} & \footnotesize{$28+21c$} & \footnotesize{$84+56c$} & $\cdots$   \\
\hline
7 & & & & & & & & \footnotesize{$1+c$} & \footnotesize{$8+7c$} & \footnotesize{$36+28c$} & $\cdots$   \\
\hline
8 & & & & & & & & & \footnotesize{$1+c$} & \footnotesize{$9+8c$} & $\cdots$   \\
\hline
9 & & & & & & & & & & \footnotesize{$1+c$} & $\cdots$   \\
\hline $\vdots$ &&&&&&&&&&& $\cdots$  \\ \hline
\end{tabular}}
\put(-16.3,97.5){\line(-3,2){11.5}}
\end{picture}

}}

\vspace{-5mm}

\begin{tabular}{lp{10.5cm}}
\textbf{FIG. 3} & Modified Pascal's triangle with boundary values
$G_{k,0} = 1$ ($k = 0,1, \ldots$) and $G_{\ell, \ell} = 1 + c$
($\ell = 1, 2, \ldots$).
\end{tabular}

\end{center}

}

\vv

To apply this result to the problem for $F_{k, \ell}^{(n)}$\,,
defined by Eq. (\ref{eq5.11}) together with Eqs. (\ref{eq5.12}) and
(\ref{eq5.13}), all we have to do is to multiply it by
$A_{2n,0}^{(n)}$ and to put $c = 1$. We get in this way: \bey
&&F_{k,\ell}^{(n)} = \left(\frac{k+\ell}{k}\right) {k
\choose \ell}A_{2n,0}^{(n)} \nonumber \\
&&(k = 2, 3, \ldots, n-1, n; \ell = 1, 2, \ldots, k-1).
\label{eq5.16} \eey

We now go through the way back from $F_{k,\ell}^{(n)}$ to
$A_{i,j}^{(n)}$. Replacing the variables $k$ and $\ell$ in Eq.
(\ref{eq5.16}) by $i$ and $j$ respectively according to Eq.
(\ref{eq5.8}), we transform $F_{k,\ell}^{(n)}$ on the left hand side
to $b_{i,j}^{(n)}$. The last-mentioned function, in the following of
Eqs. (\ref{eq5.1}) and (\ref{eq4.18}), connects to the Chebyshev
coefficient $A_{i,j}^{(n)}$ by way of the relation: \beq
A_{i,j}^{(n)} = W^{(n)}(i,j)p^{(n)}(i)b_{i,j}^{(n)} \ .
\label{eq5.17} \eeq

Using Eq. (\ref{eq4.21}) for $W^{(n)}(i,j)$ and Eq. (\ref{eq5.5})
for $p^{(n)}(i)$, the resulting expression that we obtain for
$A_{i,j}^{(n)}$ can be recast as: \bey
&&\hspace{-5mm}A_{i,j}^{(n)} = (-1)^{n-\frac{i-j}{2}}
\left[\frac{2n-1}{{2n\choose n}}\right]^2
2^{3(2n-i)} \nonumber \\
&&\hspace{-5mm} \times \frac{2j}{(i-j-1)(i+j-1)(2n-i+j)} {i+j
\choose \frac{i+j}{2}} {i-j\choose \frac{i-j}{2}} {n-\frac{i-j}{2}
\choose \frac{i+j}{2} - n} A_{2n,0}^{(n)}\ . \label{eq5.18} \eey

The range of validity of this formula is that which suffices to
cover all positions in the triangle of coefficients except the one
at the vertex on the top of the right hand side, corresponding to $i
= 2n$, $j = 0$, which we know to be occupied by the arbitrary
coefficient $A_{2n,0}^{(n)}$, and can be stated as: \beq \left.
\begin{array}{l}
i = 2n, 2n-1, 2n-2, \ldots, n+1, n; \\
j = 2n - i \ne 0, 2n-i+2, 2n-i+4, \ldots, i-2, i.
\end{array}\right\} \label{eq5.19}
\eeq

This completes the solution to the problem. In Appendices A and B we
provide the reader with tables of the multipole solutions of the
orders $n=0$ to $n = 9$ in the toroidal-polar coordinate system, in
which the overall multiplying constant for $n > 0$ was taken to be:
\beq A_{2n,0}^{(n)} = \frac{1}{2^{2n+1}}\ , \label{eq5.20} \eeq a
choice intended to bring the numbers in general to more manageable
dimensions for high values of $n$ than those of the ones afforded by
the value $A_{2n,0}^{(n)} = 1$ adopted in the main text for
illustrative purposes, while still keeping them of order unity for
low values of $n$.

Numerical computations of equilibria are usually performed in the
cylindrical system $(R, z, \phi)$ (see Fig. 1), the coordinate
$\phi$ being ignorable, and it is useful to have the multipole
solutions also expressed in this coordinate system. By defining
the normalized cylindrical coordinates as: \beq \left.
\begin{array}{l} \rho \equiv
{\dps \frac{R}{R_A}} \ \ \ \hbox{ and}  \\
\\
Z \equiv {\dps \frac{z}{R_A}}\ , \end{array}\right\} \label{eq5.21}
\eeq transformation from the toroidal-polar system can be achieved
by means of the formulae:
\beq \left. \begin{array}{l} x = \sqrt{(\rho - 1)^2 + Z^2} \ , \\
\mu = {\dps \frac{\rho - 1}{\sqrt{(\rho - 1)^2 + Z^2}}}  \ .
\end{array}\right\} \label{eq5.22} \eeq

The ensuing expressions for the multipole solutions contain only
even powers of $\rho$ and $Z$, and we thus find it convenient to
introduce the variables: \setcounter{equation}{0}
\renewcommand{\theequation}{5.22\alph{equation}}
\bey \xi &=& \rho^2 \ \ \ \hbox{ and}
\label{5.23a} \\
\nu &=& Z^2\ , \label{eq5.23b} \eey in terms of which results are
presented in Appendix C. The multipole equation they satisfy then
writes as: \setcounter{equation}{22}
\def\theequation{\thesection.\arabic{equation}}
\beq 2\xi \frac{\ptl^2\psi}{\ptl\xi^2} + 2\nu
\frac{\ptl^2\psi}{\ptl\nu^2} + \frac{\ptl\psi}{\ptl \nu} = 0\ .
\label{eq5.24} \eeq

\vv

\nd \textbf{V.1. Field lines and flux surfaces}

\vv

The discussion on the geometrical properties of the lines of force
of the magnetic fields associated with the multipole solutions is
more adequately conducted if we refer to the variables $\rho$ and
$Z$ of the cylindrical system rather than to the variables $x$ and
$\mu$ of the toroidal-polar one, both because of the greater
simplicity the expressions assume in the former system and because
of the symmetry the flux functions possess with respect to the
equatorial plane of the magnetic configurations.

Magnetic field lines lying on the flux surface associated with the
multipole solution of order $n$ are described by the equation: \beq
\vf^{(n)}(\rho, Z) = C \ , \label{eq5.25} \eeq where $C$ is a
constant. Each value of $C$ specifies a field line on a meridian
plane and the flux surface where it lies is generated by revolving
it about the axis of rotational symmetry of the configuration (the
$z$-axis). Since the quantity that ultimately bears a physical
meaning is the magnetic field rather than the flux function, any
constant can be added to the latter with no physical consequence
whatsoever and no absolute meaning  can be attached to the constant
$C$ in Eq. (\ref{eq5.25}). For the present discussion, however, it
is natural to associate the null value of the flux function with the
surface containing the ``stagnation axis'' (in the cases it does
occur), which one appears as a circumference of a circle of radius
$R_A$ and centre at $z = 0$ lying on the equatorial plane.

As general properties of the lines of force of the multipole fields
we may say that, except for the ones of the two lowest orders, they
comprise a variable number of (real) branches on each side of the
equator line (on a plane $\Phi =$ constant) dependent on the value
of $C$ in Eq. (\ref{eq5.25}), the least of these being one and the
maximal equaling the order of the multipole; that they do not close
upon themselves but extend to infinity, as it should be expected on
physical grounds, and thus that they do not encircle a point of null
field which would be identified with a magnetic axis; and that for
$C = 0$ in Eq. (\ref{eq5.25}) they become separatrices, meaning this
that they converge to or diverge from a stagnation point, which for
all of them is located at $Z = 0$, $\rho = 1$.

The multipole solution of zero order, $\vf^{(0)}(\rho, Z) = 1$,
corresponds of course to a null magnetic field. The multipole
solution of order $n=1$ is given by \beq \vf^{(1)}(\rho, Z) =
\frac{1}{4}(\rho^2 - 1)\ , \label{eq5.26} \eeq and the field lines
associated with it are described by \beq \rho = \hbox{ constant},
\label{eq5.27} \eeq which means that they are straight lines
parallel to the axis of rotational symmetry of the magnetic
configuration. This is the only case in which the magnetic field,
being uniform, vanishes at no point in space; for all other
multipole solutions, both the radial and the axial components of the
magnetic field vanish at the point $\rho = 1$, $Z = 0$, thus
justifying the designation of ``stagnation point'' given to it.

Field lines associated with the multipole solution of order $n = 2$:
\beq \vf^{(2)}(\rho, Z) = \rho^2Z^2 - \frac{1}{4}\rho^4 + \meio
\rho^2 - \frac{1}{4} \label{eq5.28} \eeq representative of the
various geometrical patterns that can be distinguished are depicted
in Fig. 4.
\begin{figure}

{\unitlength=1mm
\begin{picture}(130,95)
\put(0,95){\includegraphics[width=10cm,angle=-90]{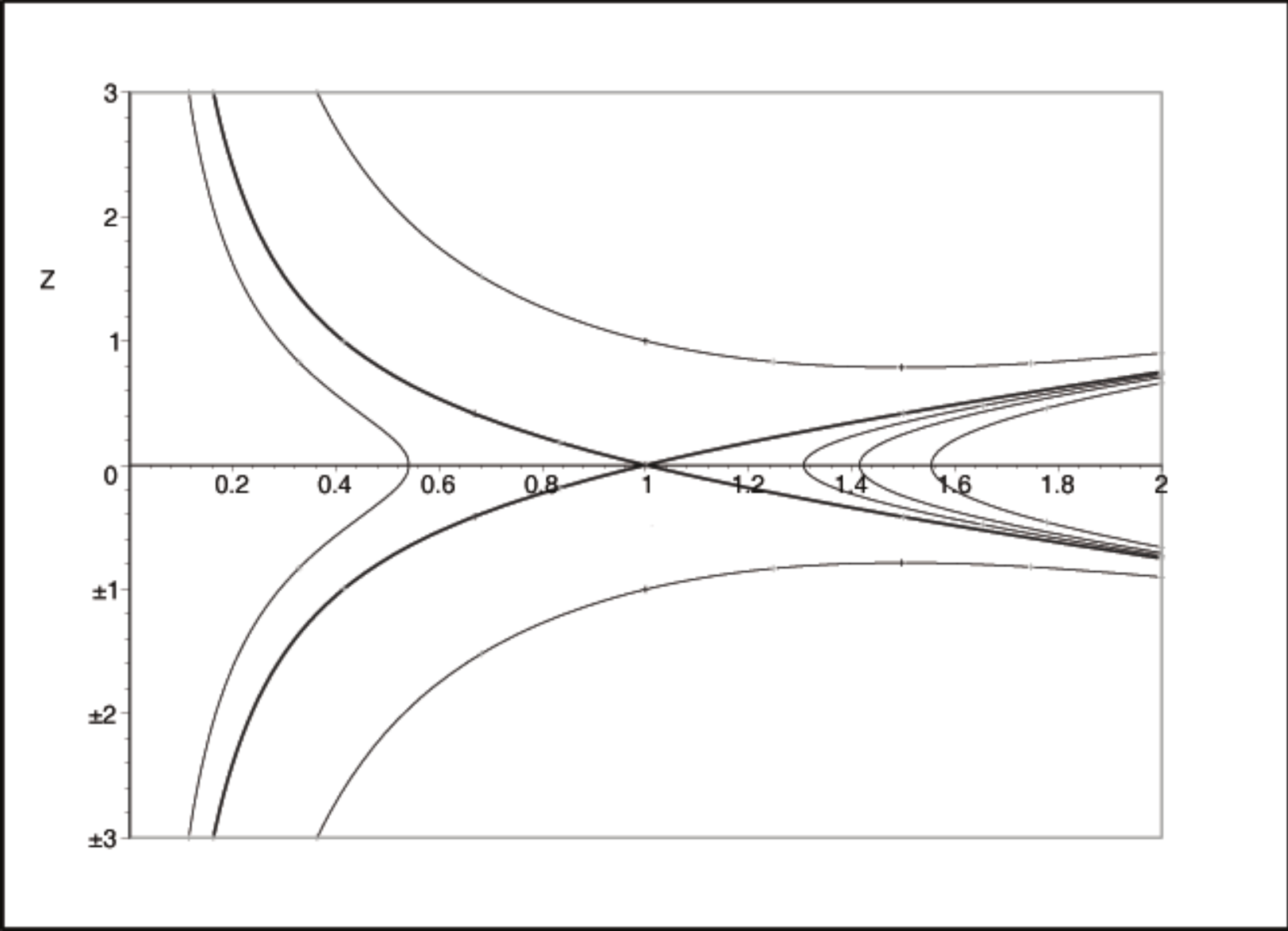}}
\put(69,38){\footnotesize $\rho$}
\end{picture}
}

\vv\vv\vv

\begin{tabular}{lp{12cm}}
{\bf FIG. 4} & Distinctive patterns assumed by the level curves of
the multipole solution of order $n=2$: $\vf^{(2)}(\rho,Z)=C$,
according to the values taken by $C$.The curve drawn in thick line
corresponds to $C=0$ and can be identified with the (trace of the)
separatrix of the multipole field (on the plane $\phi=$
constant).The curve lying in the domain of the plane $(\rho,Z)$
external to the region delimited by the branches of the separatrix
and showing two symmetrical branches with respect to the $\rho$-axis
corresponds to $C=1$. Inside the domain delimited by the branches of
the separatrix, the curve closest to its borders corresponds to
$C=-1/8$, and has two separate branches, one lying on the left of
the stagnation point $\rho=1$, $Z=0$, and the other on the right of
it. Next  to the latter it is shown the right branch of the curve
for the critical value $C=-1/4$, whose left branch coincides with
the Z-axis.Finally within the domain enclosed by the branches of the
separatrix situated on the right of the stagnation point  and the
farthest away from that point it is seen the curve for $C=-1/2$,
which has only this branch as real.
\end{tabular}

\end{figure} The field line corresponding to the value $C = 0$ for the
constant in Eq. (\ref{eq5.25}) touches the equator line $Z = 0$ at
the radial coordinate $\rho = 1$ and divides the upper and the lower
half-planes into three regions each. Looking at the upper
half-plane, the region comprised between its two branches is the
domain of the lines associated with positive values of $C$. The
field lines corresponding to negative values of $C$ with $|C| < 1/4$
are composed of two branches apart, one of which is immersed in the
region extending from the left branch of the separatrix $C = 0$ to
the axis $\rho = 0$, while the other one belongs to the region on
the right hand side of the right branch of the separatrix. The first
of the two branches coincides with the axis $\rho = 0$ as the
absolute value of $C$ is increased to $1/4$ and disappears (the
function describing it becoming complex) upon further increase of
$|C|$; the one branch that remains for negative values of $C$ with
absolute value greater than $1/4$ keeps still within the limits of
the same region containing the branches for smaller values of $|C|$,
on the right hand side of the right branch of the separatrix, and is
further removed to the right as $|C|$ is increased.

For the multipole solution of order $n = 3$, which in cylindrical
coordinates is expressed as: \beq \vf^{(3)}(\rho,Z) = \rho^2Z^4 -
\frac{3}{2}\rho^2(\rho^2 - 1)Z^2 + \frac{1}{8}(\rho^2 - 1)^3 \ ,
\label{eq5.29} \eeq Fig. 5 shows the several zones into which the
meridian plane is divided by the branches of the separatrix and by
the two branches of the field line corresponding to the critical
value $C = -1/8$ of the constant in Eq. (\ref{eq5.25}). Starting
from the $Z$-axis in the upper half-plane and moving clockwise
towards the $\rho$-axis we traverse in succession:
\begin{itemize}
\item[(a)] the region on the left of the stagnation point $\rho = 1$,
$Z = 0$ comprised between the $Z$-axis and the first branch of the
separatrix to be encountered, which contains one branch of the
field lines corresponding to $C < 0$, $|C| < 1/8$;

\item[(b)] the region delimited by the first and the second
branches of the separatrix, which lodges one branch of the field
lines with $C > 0$;

\item[(c)] the region on the right of the stagnation point
comprehended between the second branch of the separatrix, the
second branch of the critical field line (the first one being
represented by the $Z$-axis) and the third branch of the
separatrix, which is the domain of the second branch of the curves
having $C < 0$, $|C| < 1/8$;

\vv\vv\vv

{\unitlength=1mm
\begin{picture}(130,95)
\put(0,95){\includegraphics[width=10cm,angle=-90]{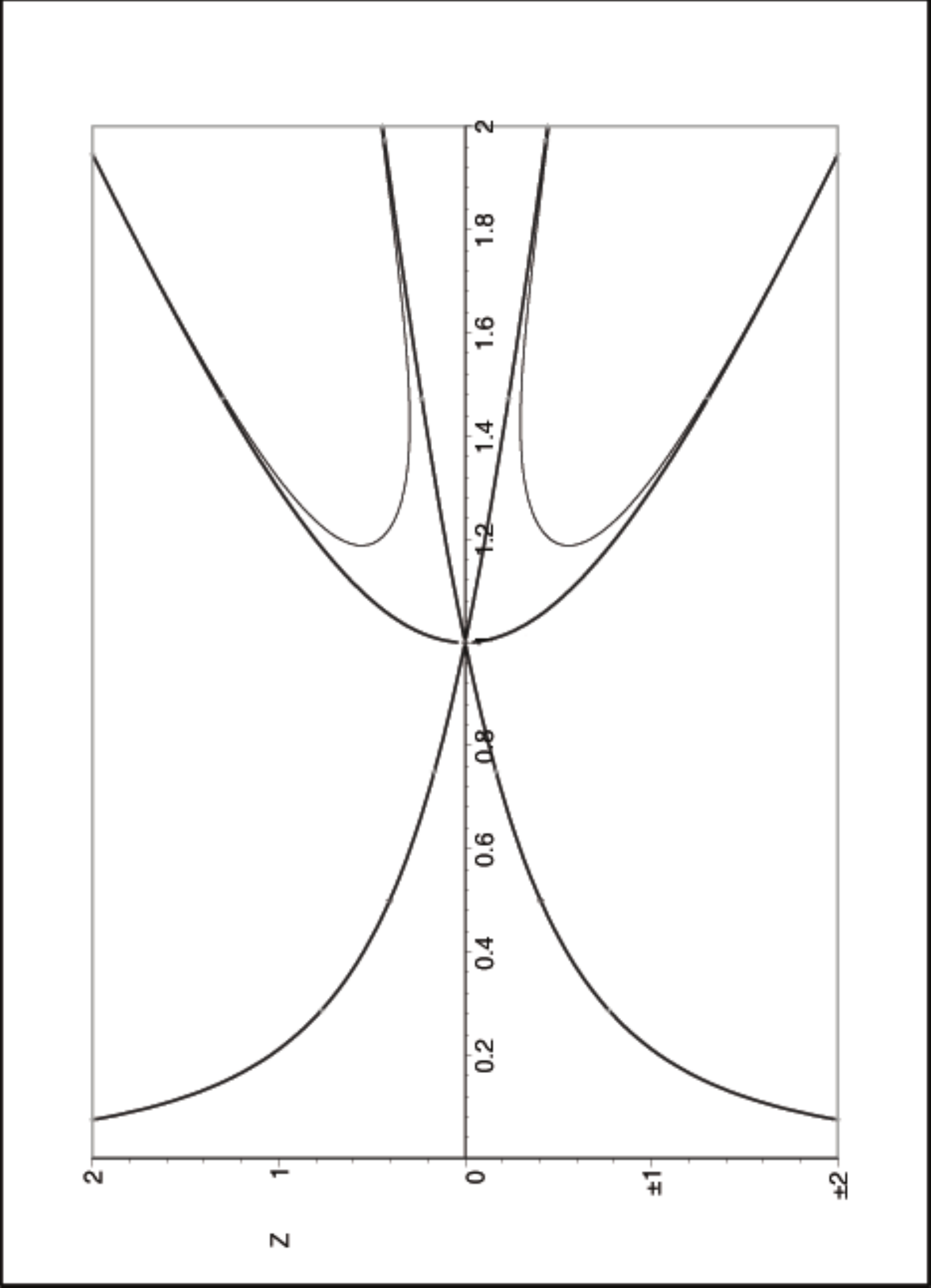}}
\put(67,39){\footnotesize $\rho$}
\end{picture}
}

\vv\vv\vv

\begin{tabular}{lp{11.7cm}}
\textbf{FIG. 5} & Particular level curves for the multipole solution
of order $n = 3$: $\vf^{(3)}(\rho,Z) = C$ on the meridian plane
$\phi =$ constant. With reference to the half-plane above the trace
of the equator plane $Z = 0$ one sees in the figure: (a) the three
branches of the separatrix, defined by $C = 0$, which are drawn in
thick line and have as common point that located at $\rho = 1$, $Z =
0$; (b) the branches of the critical field line, for which $C =
-1/8$, one of them coinciding with the $Z$-axis and the other
internal to the region delimited by the branches of the separatrix
on the right of the stagnation point.
\end{tabular}


\vv\vv

\item[(d)] the region enclosed by the second branch of the
critical field line, where the only real branch of the field lines
that have $C < 0$, $|C| > 1/8$ is immersed;

\item[(e)] the region upbounded by the third branch of the
separatrix and, within the realm of the upper half-plane, down
bounded by the $\rho$-axis, which contains the second branch of the
field lines associated with $C > 0$.
\end{itemize}

In general, a multipole solution of order $n$ gives rise to $n$
branches of the separatrix on each side of the equator line, which
divide the upper and the lower half-planes into $n+1$ zones each;
the $Z$-axis coincides with one branch of the field line associated
with the (negative) value $C_0$ of the constant $C$ obtained by
putting $\rho = 0$ in the expression of the multipole solution; the
zone limited on the left by the axis $\rho = 0$ contains one branch
of field lines defined by negative values of the constant $C$ in Eq.
(\ref{eq5.25}) whose absolute values are smaller than the absolute
value of $C_0$; going through the upper half plane clockwise we
traverse zones enclosing branches of lines of force associated with
negative values of $C$ that alternate with zones that are the
terrain of branches associated with positive values of $C$.

\vv\vv

\centerline{{\bf VI. SUMMARY}}

\setcounter{section}{6}

\vv

The general form of a multipole solution to the sourceless
Grad-Shafranov equation that is even in regard to the half-spaces
above and below the equatorial plane in the toroidal-polar
coordinate system is stated in Eq. (\ref{eq2.31}) as a polynomial in
the radial coordinate normalized to $R_A$, the distance from the
axis of rotational symmetry to the pole of the coordinate system.
The numerical value of the exponent of the lowest power in this
polynomial defines the order by which a particular multipole
solution is identified. The angular-dependent coefficient functions
of the powers of the radial coordinate variable are represented in
Eq. (\ref{eq3.26}) as combinations of Chebyshev polynomials of the
first kind having the cosine of the poloidal angle as argument. The
numerical coefficients of the Chebyshev polynomials in the
combinations can be conveniently displayed in a triangular array, an
example of which is given in Fig. 2 for the multipole solution of
order $n = 4$, and admit of being calculated by means of two
alternative sets of laws of succession, both of which require, as
starting value, that the coefficient $A_{2n,0}^{(n)}$, located at
the upper vertex on the right hand side of the triangle, be
specified arbitrarily.

The first set comprehends Eq. (\ref{eq4.1new}), Eq. (\ref{eq4.7})
and Eq. (\ref{eq4.12}), which give shape to the rules for the
sequential generation of the coefficients pertaining to the sites
forming the column on the right hand side of the array, to those
disposed along its upper side and to the remaining ones covering the
two dimensional domain on the left of the first and below the second
of these two sides, respectively.

The second set addresses to the weighted coefficients rather than to
the coefficients themselves and consists of Eq. $(4.8')$, Eq.
(\ref{eq4.25a}) in conjunction with Eq. (\ref{eq4.25b}), and Eq.
(\ref{eq4.23}) or equivalently Eq. (\ref{eq4.24}), which parallel
the equations of the first set in scope and have the same domains of
application as these. The connection between the weighted
coefficients and the coefficients proper is given by Eq.
(\ref{eq4.18}); its use requires the knowledge of the reciprocal of
the function $W^{(n)}(i,j)$, which one can be evaluated with the
help of Eq. (\ref{eq4.16}) taking unity as the initial value of the
sequence corresponding to a fixed $i$, as specified by Eq.
(\ref{eq4.19}), and recalling that the function $h(i,j)$ is defined
by Eq. (\ref{eq4.13}).

Besides being calculable by these two recursive schemes, the
coefficients can be obtained from a single expression, given by Eq.
(\ref{eq5.18}), which encompasses the solutions to the complete set
of difference equations for the Chebyshev coefficients and whose
range of validity reaches every site in the triangle of
coefficients.

Except for the multipole solution of order $n=1$, which gives a
magnetic field constant and parallel to the axis of rotational
symmetry, the multipole fields in general vanish at the radial
coordinate $r = 0$.

Tables of the even multipole solutions of order $n = 0$ to $n = 9$
in variables of the toroidal-polar coordinate system are provided in
Appendices A and B, and in variables of the cylindrical coordinate
system in Appendix C.


\vv\vv

\begin{center}
{\bf APPENDIX A: EXPRESSIONS FOR THE EVEN MULTIPOLE SOLUTIONS OF
ORDERS $n = 0$ TO $n=9$ IN THE TOROIDAL-POLAR COORDINATE SYSTEM IN
TERMS OF THE HARMONICS OF THE POLAR ANGLE $\ta$}
\end{center}

\setcounter{equation}{0}
\def\theequation{A.\arabic{equation}}

\vv

The variables are: $x = r / R_A$, where $r$ is the radial coordinate
on the meridian plane and $R_A$ is the distance from the axis of
rotational symmetry to the pole of the coordinate system measured on
the meridian plane; and $\theta$, the polar angle on the meridian
plane.

The multipole solutions, when their angular dependences are
expressed in terms of the harmonics of the polar angle $\ta$, can be
stated in the following general form: \beq \vf^{(n)}(x,\ta) =
\sum_{j=0}^{2n} M_{jn}(x) \cos j\ta\ \ \ (n = 1, 2, 3, \ldots)\ ,
\label{eqa1} \eeq where the coefficients $M_{jn}(x)$ are given by:
\begin{itemize}
\item[(a)] for $j = 0, 1, 2, \ldots, n$,
\beq M_{jn}(x) = \sum_{l=0}^{L_1} A_{2n-j+2l, j}^{(n)}x^{2n-j+2l}
\label{eqa2} \eeq with \beq L_1 = \left\{\begin{array}{ll} {\dps
\frac{j}{2}}&\hbox{ for } j = 0 \hbox{ or even} \\
& \\
{\dps \frac{j-1}{2}} &\hbox{ for } j \hbox{ odd;}
\end{array}\right. \label{eqa3}
\eeq

\item[(b)] for $j = n+1, n+2, \ldots, 2n$,
\beq M_{jn}(x) = \sum_{l=0}^{L_2} A_{2j+2l, j}^{(n)}x^{j+2l}
\label{eqa4} \eeq with \beq L_2 = \left\{\begin{array}{ll} {\dps
\frac{2n -j}{2}}&\hbox{ for } j \hbox{ even} \\
& \\
{\dps \frac{2n-1-j}{2}} &\hbox{ for } j \hbox{ odd.}
\end{array}\right. \label{eqa5}
\eeq

\end{itemize}

The Chebychev coefficients $A_{k,j}^{(n)}$ are evaluated by means of
the formula in Eq. (\ref{eq5.18}) or by the recursive rules stated
in Section IV.

The multipole solutions referred to in the title of this Appendix
are listed below.
\bey \vf^{(0)}(x, \theta )&=&1\ , \label{eqa6} \\
&& \nonumber \\
\vf^{(1)}(x, \theta )&=&{\displaystyle \frac {1}{8}} x^{2}
\mathrm{cos}(2 \theta ) + {\displaystyle \frac {1}{2}}
\mathrm{cos}(\theta ) x  + {\displaystyle \frac {x^{2}}{8}}\ ,
\label{eqa7} \\
&& \nonumber \\
 \vf^{( 2)}(x,  \theta )&=&
 - {\displaystyle \frac {5}{32}}  \mathrm{cos}(4 \theta ) x^{4
} - {\displaystyle \frac {3}{4}}  \mathrm{cos}(3 \theta ) x^{3} + (
- {\displaystyle \frac {1}{8}}  x^{4} - x^{2 }) \mathrm{cos}(2
\theta ) \nonumber \\
&&- {\displaystyle \frac {1}{4} }
 \mathrm{cos}(\theta ) x^{3} + {\displaystyle \frac {x^{4}
}{32}} \ , \label{eqa8} \\
&& \nonumber \\
 \vf^{( 3)}(x,  \theta )&=&{\displaystyle \frac
{21}{256}}
 \mathrm{ cos}(6 \theta ) x^{6} + {\displaystyle \frac
{35}{64}}  \mathrm{cos}(5 \theta ) x^{5} + ({\displaystyle \frac
{7}{128}}  x^{6} + {\displaystyle \frac {5}{4}}  x^{4})
\mathrm{cos}(4 \theta ) \nonumber \\
&& + ({\displaystyle \frac
{15}{64}} x^{5} + x^{3}) \mathrm{cos}(3 \theta ) + ( -
{\displaystyle \frac {5}{ 256}}
 x^{6} + {\displaystyle \frac {1}{4}}  x^{4}) \mathrm{
cos}(2 \theta ) \nonumber \\
&&- {\displaystyle \frac {1}{32}} \mathrm{cos}(\theta)
x^{5}  + {\displaystyle \frac {x^{6}}{ 128}} \ , \label{eqa9} \\
&& \nonumber \\
\vf^{( 4)}(x,  \theta )&=& - {\displaystyle \frac {429}{12800}}
\mathrm{cos}(8 \theta ) x^{8}  - {\displaystyle \frac { 231}{800}}
\mathrm{cos}(7 \theta ) x^{7} + ( - {\displaystyle \frac {33}{1600}}
x^{8} - {\displaystyle \frac {
189}{200}}  x^{6}) \mathrm{cos}(6 \theta ) \nonumber \\
&& + ( - {\displaystyle \frac {21}{160}}  x^{7} - {\displaystyle
\frac {7}{5}}  x^{5}) \mathrm{cos}(5 \theta ) + ({\displaystyle
\frac {21}{3200}}  x^{8} - {\displaystyle \frac {7}{25}}  x^{6} -
{\displaystyle \frac {4}{
5}}  x^{4}) \mathrm{cos}(4 \theta ) \nonumber \\
&& + ({\displaystyle \frac {21}{800}}  x^{7} - {\displaystyle \frac
{1}{5}}  x^{5}) \mathrm{cos}(3 \theta ) + ( - {\displaystyle \frac
{7}{1600}}  x^{8} + {\displaystyle \frac {1}{40}}  x^{6})
\mathrm{cos}(2 \theta ) \nonumber \\
&&- {\displaystyle \frac {1}{160}} \mathrm{cos}(\theta ) x^{7} +
{\displaystyle \frac {x^{8}}{512}}\ ,
\label{eqa10} \\
&& \nonumber \\
\vf^{( 5)}(x,  \theta )&=&{\displaystyle \frac {2431}{200704}}
\mathrm{cos}(10 \theta ) x^{10} + {\displaystyle
\frac {6435}{50176}}  \mathrm{cos}(9 \theta ) x^{9} \nonumber \\
&&+ ({\displaystyle \frac {715}{100352}}  x^{10} + {\displaystyle
\frac {429}{784}}  x^{8}) \mathrm{cos}(8 \theta
 ) + ({\displaystyle \frac {429}{7168}}  x^{9} +
{\displaystyle \frac {33}{28}}  x^{7}) \mathrm{cos}(7 \theta )
 \nonumber \\
&& + ( - {\displaystyle \frac {429}{200704}}  x^{10} +
{\displaystyle \frac {297}{1568}}  x^{8} + {\displaystyle \frac
{9}{7}}  x^{6}) \mathrm{cos}(6 \theta ) \nonumber \\
&&+ ( - {\displaystyle \frac {165}{12544}}  x^{9} + {\displaystyle
\frac {15}{56}}  x^{7} + {\displaystyle \frac {4 }{7}}  x^{5})
\mathrm{cos}(5 \theta ) \nonumber \\
&& + ( {\displaystyle \frac {33}{25088}} x^{10} - {\displaystyle
\frac {3}{112}}  x^{8} + {\displaystyle \frac {1}{7}} x^{6})
\mathrm{cos}(4 \theta ) + ({\displaystyle \frac {9}{1792}} x^{9} -
{\displaystyle \frac {1}{56}}  x^{7}) \mathrm{cos}(3 \theta )
\nonumber \\
&& + ( - {\displaystyle \frac {15}{14336}}  x^{10} + {\displaystyle
\frac {1}{224}}  x^{8}) \mathrm{cos}(2 \theta )
 - {\displaystyle \frac {5}{3584}}  \mathrm{cos}(\theta )
 x^{9} + {\displaystyle \frac {x^{10}}{2048}} \ , \label{eqa11} \\
 && \nonumber \\
\vf^{( 6)}(x,  \theta )&=& - {\displaystyle \frac {4199}{1032192}}
\mathrm{cos}(12 \theta ) x^{12}  - {\displaystyle \frac
{46189}{903168}}  \mathrm{cos}(11 \theta ) x^{11} \nonumber \\
&& + ( - {\displaystyle \frac {4199}{1806336}}  x^{12} -
{\displaystyle \frac {60775}{225792}}  x^{10}) \mathrm{cos}(10
 \theta ) \nonumber \\
 &&+ ( - {\displaystyle \frac {2431}{100352}}  x
^{11} - {\displaystyle \frac {3575}{4704}}  x^{9}) \mathrm{cos}
(9 \theta ) \nonumber \\
&& + ({\displaystyle \frac {2431}{3612672}}  x^{12} - {\displaystyle
\frac {715}{7056}}  x^{10} - {\displaystyle \frac {715}{588}} x^{8})
\mathrm{cos}(8 \theta ) \nonumber \\
&&+ ({\displaystyle \frac {715}{129024}} x^{11} - {\displaystyle
\frac {143}{672}} x^{9} - {\displaystyle
\frac {22}{21}}  x^{7}) \mathrm{cos}(7 \theta ) \nonumber \\
&&+ ( - {\displaystyle \frac {715}{1806336}}  x^{12} +
{\displaystyle \frac {429}{25088}}  x^{10} - {\displaystyle \frac
{11}{49}}  x^{8} - {\displaystyle \frac {8}{21}}  x^{6})
 \mathrm{cos}(6 \theta ) \nonumber \\
&& + ( - {\displaystyle \frac {715}{301056}}  x^{11} +
{\displaystyle \frac {55}{2352}}  x^{9} - {\displaystyle \frac {
2}{21}}  x^{7}) \mathrm{cos}(5 \theta ) \nonumber \\
&&+ ({\displaystyle \frac {715}{2408448}}  x^{12} - {\displaystyle
\frac {11}{2352}} x^{10} + {\displaystyle \frac {1}{84}}  x^{8})
\mathrm{cos}(4 \theta )\nonumber \\
&& + ( {\displaystyle \frac {55}{50176}}  x^{11} - {\displaystyle
\frac {1}{336}}  x^{9}) \mathrm{cos}(3 \theta ) + ( - {\displaystyle
\frac {11}{43008}}  x^{12} + {\displaystyle \frac {5}{5376}} x^{10})
\mathrm{cos}(2 \theta ) \nonumber \\
&&- {\displaystyle \frac {1}{3072}} \mathrm{cos}(\theta
 ) x^{11} + {\displaystyle \frac {x^{12}}{8192}} \ , \label{eqa12} \\
 && \nonumber \\
\vf^{( 7)}(x,  \theta )&=&{\displaystyle \frac {185725}{142737408}}
\mathrm{cos}(14 \theta ) x^{14} + {\displaystyle \frac
{676039}{35684352}}  \mathrm{cos}(13 \theta )x^{13}\nonumber \\
&& + ({\displaystyle \frac {52003}{71368704}}  x^{14} +
{\displaystyle \frac {29393}{247808}}  x^{12}) \mathrm{cos}(12
 \theta ) \nonumber \\
 &&+ ({\displaystyle \frac {29393}{3244032}}  x^{13} +
{\displaystyle \frac {20995}{50688}}  x^{11}) \mathrm{cos}(11
\theta ) \nonumber \\
&& + ( - {\displaystyle \frac {29393}{142737408}}  x^{14}
 + {\displaystyle \frac {104975}{2230272}}  x^{12} +
{\displaystyle \frac {5525}{6336}}  x^{10}) \mathrm{cos}(10
\theta ) \nonumber  \\
&& + ( - {\displaystyle \frac {4199}{1982464}}  x^{13} +
{\displaystyle \frac {1105}{8448}}  x^{11} + {\displaystyle
\frac {195}{176}}  x^{9}) \mathrm{cos}(9 \theta ) \nonumber \\
&& + ({\displaystyle \frac {4199}{35684352}}  x^{14} -
{\displaystyle \frac {221}{25344}}  x^{12} + {\displaystyle \frac
{325}{1584}}  x^{10} + {\displaystyle \frac {26}{33}}  x
^{8}) \mathrm{cos}(8 \theta ) \nonumber \\
&& + ({\displaystyle \frac {1547}{1622016}}  x^{13} - {\displaystyle
\frac {455}{25344}}  x^{11} + {\displaystyle \frac {91}{528}}  x^{9}
+ {\displaystyle \frac {8}{33}}  x^{7})
 \mathrm{cos}(7 \theta ) \nonumber \\
&& + ( - {\displaystyle \frac {1105}{12976128}}  x^{14} +
{\displaystyle \frac {65}{22528}}  x^{12} - {\displaystyle \frac
{13}{704}}  x^{10} + {\displaystyle \frac {2}{33}}  x^{8}
) \mathrm{cos}(6 \theta ) \nonumber \\
&& + ( - {\displaystyle \frac {1625}{3244032}}  x^{13} +
{\displaystyle \frac {65}{16896}}  x^{11} - {\displaystyle
\frac {1}{132}}  x^{9}) \mathrm{cos}(5 \theta ) \nonumber \\
&& + ({\displaystyle \frac {455}{6488064}}  x^{14} - {\displaystyle
\frac {65}{67584}}  x^{12} + {\displaystyle \frac {1}{528}}  x^{10})
\mathrm{cos}(4 \theta ) \nonumber \\
&&+ ({\displaystyle \frac {91}{360448}} x^{13} - {\displaystyle
\frac {5}{8448}}  x^{11}) \mathrm{cos}(3 \theta
 ) \nonumber \\
 && + ( - {\displaystyle \frac {91}{1441792}}  x^{14} +
{\displaystyle \frac {7}{33792}}  x^{12}) \mathrm{cos}(2 \theta )
\nonumber \\
&&- {\displaystyle \frac {7}{90112}}  \mathrm{cos}(\theta ) x^{13} +
{\displaystyle \frac {x^{14}}{32768}}\ , \label{eqa13} \\
&& \nonumber \\
\vf^{( 8)}(x,  \theta )&=& - {\displaystyle \frac
{1077205}{2680291328 }} \mathrm{cos}(16 \theta ) x^{16} -
{\displaystyle \frac {557175}{83759104}}  \mathrm{cos}(15 \theta )
x^{15} \nonumber \\
&& + ( - {\displaystyle \frac {37145}{167518208}}  x^{16}
 - {\displaystyle \frac {9100525}{188457984}}  x^{14}) \mathrm{
cos}(14 \theta ) \nonumber \\
&&+ ( - {\displaystyle \frac {185725}{57987072}} x^{15}
 - {\displaystyle \frac {364021}{1812096}}  x^{13}) \mathrm{cos
}(13 \theta ) \nonumber \\
&& + ({\displaystyle \frac {185725}{3015327744}}  x^{16} -
{\displaystyle \frac {52003}{2617472}}  x^{14} - {\displaystyle
\frac {79135}{151008}}  x^{12}) \mathrm{cos}(12 \theta ) \nonumber
\\
&&+ ({\displaystyle \frac {52003}{68530176}}  x^{15} -
{\displaystyle \frac {11305}{164736}}  x^{13} - {\displaystyle \frac
{2261}{2574}} x^{11}) \mathrm{cos}(11 \theta ) \nonumber \\
&& + ( - {\displaystyle \frac {52003}{1507663872}}  x^{16}
 + {\displaystyle \frac {56525}{14496768}}  x^{14} -
{\displaystyle \frac {8075}{56628}}  x^{12} - {\displaystyle \frac
{1190}{1287}}  x^{10}) \mathrm{cos}(10 \theta ) \nonumber \\
&&+ ( - {\displaystyle \frac {2261}{6443008}}  x^{15} +
{\displaystyle \frac {1615}{151008}}  x^{13} - {\displaystyle \frac
{51}{286}} x^{11} - {\displaystyle \frac {80}{143}}  x^{
9}) \mathrm{cos}(9 \theta ) \nonumber \\
&& + ({\displaystyle \frac {11305}{463896576}}  x^{16} \!-\!
{\displaystyle \frac {323}{226512}}  x^{14} \!+\! {\displaystyle
\frac {85}{5148}}  x^{12} \!-\! {\displaystyle \frac {160}{1287}}
x^{10} - {\displaystyle \frac {64}{429}}  x^{8}) \mathrm{cos}(8
 \theta ) \nonumber \\
&& + ({\displaystyle \frac {11305}{57987072}}  x^{15} -
{\displaystyle \frac {119}{41184}}  x^{13} + {\displaystyle \frac
{35}{2574}}  x^{11} - {\displaystyle \frac {16}{429}}  x
^{9}) \mathrm{cos}(7 \theta ) \nonumber \\
&& + ( - {\displaystyle \frac {2261}{115974144}}  x^{16} +
{\displaystyle \frac {85}{146432}}  x^{14} - {\displaystyle \frac
{5}{1716}}  x^{12} + {\displaystyle \frac {2}{429}}  x^{
10}) \mathrm{cos}(6 \theta ) \nonumber \\
&& + ( - {\displaystyle \frac {595}{5271552}}  x^{15} +
{\displaystyle \frac {125}{164736}}  x^{13} - {\displaystyle
\frac {1}{858}}  x^{11}) \mathrm{cos}(5 \theta ) \nonumber \\
&& + ({\displaystyle \frac {119}{7028736}}  x^{16} - {\displaystyle
\frac {35}{164736}}  x^{14} + {\displaystyle
\frac {5}{13728}}  x^{12}) \mathrm{cos}(4 \theta ) \nonumber \\
&&+ ({\displaystyle \frac {35}{585728}}  x^{15} - {\displaystyle
\frac {7}{54912}}  x^{13}) \mathrm{cos}(3  \theta ) \nonumber \\
&&+\! (\! - {\displaystyle \frac {5}{319488}}  x^{16}\!\! +\!
{\displaystyle \frac {7}{146432}}  x^{14}) \mathrm{cos}(2 \theta )
\!- \!{\displaystyle \frac {1}{53248}} \mathrm{cos}(\theta ) x^{15}
\!\!+\! {\displaystyle \frac {x^{16}}{131072}}, \label{eqa14} \\
&& \nonumber \\
\vf^{( 9)}(x, \theta )&=&{\displaystyle \frac {1178589}{9746513920}}
\mathrm{cos}(18 \theta ) x^{18} + {\displaystyle \frac
{60108039}{26802913280}}
 \mathrm{cos}(17 \theta ) x^{17} \nonumber \\
&& + ({\displaystyle \frac {3535767}{53605826560}}  x^{18}
 + {\displaystyle \frac {1938969}{104698880}}  x^{16}) \mathrm{
cos}(16 \theta ) \nonumber \\
&& + ({\displaystyle \frac {5816907}{5360582656}}  x^{17}
 + {\displaystyle \frac {468027}{5234944}}  x^{15}) \mathrm{cos
}(15 \theta ) \nonumber \\
&& + ( - {\displaystyle \frac {1938969}{107211653120}}  x^{ 18} +
{\displaystyle \frac {3276189}{418795520}}  x^{16} + {\displaystyle
\frac {364021}{1308736}}  x^{14}) \mathrm{cos}(
14 \theta ) \nonumber \\
&& + ( - {\displaystyle \frac {66861}{257720320}}  x^{17}
 + {\displaystyle \frac {52003}{1610752}}  x^{15} +
{\displaystyle \frac {364021}{629200}}  x^{13}) \mathrm{cos}(13
 \theta ) \nonumber  \\
&& + ({\displaystyle \frac {66861}{6700728320}}  x^{18}\!\! -\!
{\displaystyle \frac {66861}{41879552}}  x^{16} \!\!+\!
{\displaystyle \frac {1092063}{13087360}}  x^{14} \!\!+\!
{\displaystyle \frac {31654}{39325}}  x^{12}) \mathrm{cos}(12
\theta ) \nonumber \\
&& + ({\displaystyle \frac {7429}{60915712}}  x^{17} -
{\displaystyle \frac {52003}{9518080}}  x^{15} + {\displaystyle
\frac {15827}{114400}}  x^{13} + {\displaystyle \frac {2584}{
3575}}  x^{11}) \mathrm{cos}(11 \theta ) \nonumber  \\
&&+ ( - {\displaystyle \frac {37145}{5360582656}}  x^{18} +
{\displaystyle \frac {52003}{83759104}}  x^{16} - {\displaystyle
\frac {2261}{201344}}  x^{14} \nonumber \\
&&+ {\displaystyle \frac {2261}{15730}}
 x^{12} + {\displaystyle \frac {272}{715}}
 x^{10}) \mathrm{cos}(10 \theta ) \nonumber \\
&& + ( - {\displaystyle \frac {468027}{6700728320}}  x^{17}
 + {\displaystyle \frac {6783}{4026880}}  x^{15} -
{\displaystyle \frac {8721}{629200}}  x^{13} \nonumber \\
&&+ {\displaystyle \frac {306}{3575}}  x^{11} + {\displaystyle \frac
{64}{715}}  x
^{9}) \mathrm{cos}(9 \theta ) \nonumber  \\
&& + ({\displaystyle \frac {364021}{67007283200}}  x^{18}
 - {\displaystyle \frac {2261}{8053760}}  x^{16} +
{\displaystyle \frac {323}{125840}}  x^{14} \nonumber \\
&&- {\displaystyle \frac {34}{3575}}  x^{12} + {\displaystyle \frac
{16}{715}}  x
^{10}) \mathrm{cos}(8 \theta ) \nonumber \\
&& + ({\displaystyle \frac {110789}{2577203200}}  x^{17} -
{\displaystyle \frac {2261}{4026880}}  x^{15} + {\displaystyle \frac
{119}{57200}}  x^{13} - {\displaystyle \frac {2}{715}}  x
^{11}) \mathrm{cos}(7 \theta ) \nonumber \\
&& + ( - {\displaystyle \frac {47481}{10308812800}}  x^{18}
 + {\displaystyle \frac {20349}{161075200}}  x^{16} -
{\displaystyle \frac {51}{91520}}  x^{14} + {\displaystyle
\frac {1}{1430}}  x^{12}) \mathrm{cos}(6 \theta ) \nonumber \\
&& + ( - {\displaystyle \frac {6783}{257720320}}  x^{17} +
{\displaystyle \frac {119}{732160}}  x^{15} - {\displaystyle
\frac {1}{4576}}  x^{13}) \mathrm{cos}(5 \theta ) \nonumber \\
&& + ({\displaystyle \frac {969}{234291200}}  x^{18} -
{\displaystyle \frac {357}{7321600}}  x^{16} + {\displaystyle
\frac {7}{91520}}  x^{14}) \mathrm{cos}(4 \theta ) \nonumber \\
&& + ({\displaystyle \frac {153}{10649600}}  x^{17} - {\displaystyle
\frac {21}{732160}}  x^{15}) \mathrm{cos}(3
\theta ) \nonumber \\
&& + ( - {\displaystyle \frac {51}{13107200}}  x^{18} +
{\displaystyle \frac {3}{266240}}  x^{16}) \mathrm{cos}(2  \theta )
 \nonumber \\
 &&- {\displaystyle \frac {3}{655360}}  \mathrm{cos }(\theta ) x^{17}
+ {\displaystyle \frac {x^{18}}{524288}}\ . \label{eqa15} \eey

\vv\vv

\begin{center}
{\bf APPENDIX B: EXPRESSIONS FOR THE EVEN MULTIPOLE SOLUTIONS OF
ORDERS $n = 0$ TO $n=9$ IN THE TOROIDAL-POLAR COORDINATE SYSTEM AS
BIVARIATE POLYNOMIALS IN THE NORMALIZED RADIAL VARIABLE AND IN THE
COSINE OF THE POLOIDAL ANGLE}
\end{center}

\setcounter{equation}{-1}
\def\theequation{B.\arabic{equation}}

\vv

The variables are: $x$, as defined in Appendix A, and $\mu \equiv
\cos\ta$.
\bey &&\hspace{-7mm}\vf^{(0)}(x, \mu) = 1\ , \label{eqb0} \\
&&\hspace{-7mm} \nonumber \\
&&\hspace{-7mm}\vf^{(1)}(x,\mu) = \frac{1}{2}\mu x + \frac{1}{4}\mu^2x^2 \ , \label{eqb1} \\
&&\hspace{-7mm} \nonumber \\
&&\hspace{-7mm}\vf^{(2)} (x,\mu) = ( - {\displaystyle \frac {5}{4}}
\mu ^{4} + \mu ^{2})  x^{4} + ( - 3 \mu ^{3} + 2 \mu ) x^{3} + ( - 2
\mu ^{2} + 1
) x^{2} \ , \label{eqb2} \\
&&\hspace{-7mm} \nonumber \\
&&\hspace{-7mm}\vf^{(3)}(x,\mu) = ({\displaystyle \frac {21}{8}}
 \mu ^{6} - {\displaystyle \frac {7}{2}}  \mu ^{4} + \mu
^{2}) x^{6} + ( {\displaystyle \frac {35}{4}}  \mu ^{5} - 10 \mu
^{3} + 2 \mu
) x^{5}  \nonumber \\
&&\hspace{-7mm}\quad + (10 \mu ^{4} - {\displaystyle \frac {19}{2}}
 \mu ^{ 2} + 1) x^{4}
+ (4 \mu ^{3} - 3 \mu ) x^{3} \ , \label{eqb3} \\
&&\hspace{-7mm} \nonumber \\
&&\hspace{-7mm}\vf^{(4)}(x,\mu) = ( \!-\! {\displaystyle \frac
{429}{100}} \mu ^{8 } \!+\! {\displaystyle \frac {198}{25}} \mu
^{6} \!-\! {\displaystyle \frac {108}{25}} \mu ^{4} \!+\!
{\displaystyle \frac {16}{25}}\mu ^{2})x^{8} \!+\! ( \!-\!
{\displaystyle \frac {462}{25}} \mu ^{7} \! +\! {\displaystyle
\frac {756}{25}}\mu ^{5} \!-\! {\displaystyle \frac {336}{25}} \mu
^{3} \!+\! {\displaystyle \frac {32}{25}} \mu )x^{7} \nonumber \\
&&\hspace{-7mm} \quad +\! ( - {\displaystyle \frac {756}{25}}
 \mu ^{6} + {\displaystyle \frac {1078}{25}}  \mu ^{4} -
{\displaystyle \frac {368}{25}}  \mu ^{2} + {\displaystyle \frac
{16}{25}} ) x ^{6} + ( - {\displaystyle \frac {112}{5}}  \mu ^{5} +
{\displaystyle \frac {136}{5}}  \mu ^{3} - {\displaystyle \frac
{32}{5}}  \mu ) x^{5}
\nonumber \\
&&\hspace{-7mm}\quad  + ( - {\displaystyle \frac {32}{5}}  \mu ^{4}
+ {\displaystyle \frac {32}{5}}  \mu ^{2} - {\displaystyle
\frac {4}{5}} ) x^{4} \ , \label{eqb4} \\
&&\hspace{-7mm} \nonumber \\
&&\hspace{-7mm} \vf^{(5)}(x,\mu) = ({\displaystyle \frac
{2431}{392}}  \mu ^{10}
 - {\displaystyle \frac {715}{49}}  \mu ^{8} + {\displaystyle
\frac {572}{49}}  \mu ^{6} - {\displaystyle \frac {176}{49}}
\mu ^{4} + {\displaystyle \frac {16}{49}}  \mu ^{2}) x^{10} \nonumber \\
&&\hspace{-7mm}\quad   + ({\displaystyle \frac {6435}{196}}  \mu
^{9} - {\displaystyle \frac {3432}{49}}  \mu ^{7} + {\displaystyle
\frac {2376}{49}}  \mu ^{5} - {\displaystyle \frac {576}{49}}
\mu ^{3} + {\displaystyle \frac {32}{49}}  \mu ) x^{9} \nonumber \\
 &&\hspace{-7mm}\quad    + ({\displaystyle \frac {3432}{49}}  \mu ^{8} -
{\displaystyle \frac {6567}{49}}  \mu ^{6} + {\displaystyle \frac
{3834}{49}}  \mu ^{4} - {\displaystyle \frac {680}{49}}
\mu ^{2} + {\displaystyle \frac {16}{49}} ) x^{8} \nonumber \\
 &&\hspace{-7mm}\quad    + ({\displaystyle \frac {528}{7}}  \mu ^{7} -
{\displaystyle \frac {894}{7}}  \mu ^{5} + {\displaystyle \frac
{424}{7}}  \mu ^{3} - {\displaystyle \frac {48}{7}}  \mu ) x^{7} +
({\displaystyle \frac {288}{7}}  \mu ^{6} - {\displaystyle \frac
{424}{7}}  \mu ^{4} + 22 \mu ^{2} -
{\displaystyle \frac {8}{7}} ) x^{6} \nonumber \\
&&\hspace{-7mm}\quad + ({\displaystyle \frac {64}{7}}  \mu ^{5} -
{\displaystyle \frac {80}{7}}  \mu ^{3} + {\displaystyle \frac {
20}{7}}  \mu ) x^{5} \ ,  \label{eqb5} \\
&&\hspace{-7mm} \nonumber \\
&&\hspace{-7mm} \vf^{(6)}(x,\mu) = ( - {\displaystyle \frac
{4199}{504}} \mu ^{ 12} + {\displaystyle \frac {20995}{882}} \mu
^{10} - {\displaystyle \frac {11050}{441}} \mu ^{8} +
{\displaystyle \frac {5200}{441}}\mu ^{6} - {\displaystyle \frac
{1040}{441} } \mu ^{4} + {\displaystyle \frac {64}{441}}
\mu ^{2})x^{12} \nonumber \\
 &&\hspace{-7mm}\quad   + ( - {\displaystyle \frac {46189}{882}}  \mu ^{11}
 + {\displaystyle \frac {60775}{441}}  \mu ^{9} -
{\displaystyle \frac {57200}{441}}  \mu ^{7} + {\displaystyle \frac
{22880}{441}}  \mu ^{5} - {\displaystyle \frac {3520}{441}
}  \mu ^{3} + {\displaystyle \frac {128}{441}}  \mu ) x^{11} \nonumber \\
 &&\hspace{-7mm}\quad   + ( - {\displaystyle \frac {60775}{441}}  \mu ^{10}
 + {\displaystyle \frac {292435}{882}}  \mu ^{8} -
{\displaystyle \frac {121264}{441}}  \mu ^{6} + {\displaystyle \frac
{39952}{441}}  \mu ^{4} - {\displaystyle \frac {4352}{441}
}  \mu ^{2} + {\displaystyle \frac {64}{441}} ) x^{10} \nonumber \\
 &&\hspace{-7mm}\quad    + ( - {\displaystyle \frac {28600}{147}}  \mu ^{9}
 + {\displaystyle \frac {62348}{147}}  \mu ^{7} -
{\displaystyle \frac {44704}{147}}  \mu ^{5} + {\displaystyle \frac
{11584}{147}}  \mu ^{3} - {\displaystyle \frac {256}{49}}
 \mu ) x^{9} \nonumber \\
 &&\hspace{-7mm}\quad    + ( - {\displaystyle \frac {22880}{147}}  \mu ^{8}
 + {\displaystyle \frac {44704}{147}}  \mu ^{6} -
{\displaystyle \frac {27002}{147}}  \mu ^{4} + {\displaystyle \frac
{1704}{49}}  \mu ^{2} - {\displaystyle \frac {48}{49}} )
x^{8} \nonumber \\
 &&\hspace{-7mm}\quad    + \!( \!- {\displaystyle \frac {1408}{21}} \mu ^{7}
 \!+\!{\displaystyle \frac {2432}{21}} \mu ^{5} \!- {\displaystyle \frac
{1192}{21}} \mu ^{3} \!+ \!{\displaystyle \frac {48}{7}}  \mu
)x^{7} \!+\! ( \!- {\displaystyle \frac {256}{21}} \mu ^{6} \!+\!
{\displaystyle \frac {128}{7}}\mu ^{4} \!- {\displaystyle \frac
{48}{7}} \mu ^{2} \!+\! {\displaystyle \frac {8}{21}} )x^{6
}   , \label{eqb6} \\
&&\hspace{-7mm} \nonumber \\
&&\hspace{-7mm}\vf^{(7)}\!(x,\mu)\!= \!({\displaystyle \frac
{185725}{17424}} \mu ^{14} \!\!\!-\! {\displaystyle \frac
{52003}{1452}} \mu ^{12} \!\!+\! {\displaystyle \frac
{11305}{242}} \mu ^{10} \!\!\!-\! {\displaystyle \frac
{32300}{1089}} \mu ^{8} \!\!+\! {\displaystyle \frac {3400}{363 }}
\mu ^{6} \!\!\!-\! {\displaystyle \frac {160}{121}}
\mu ^{4} \!\!+\!{\displaystyle \frac {64}{1089}} \mu ^{2}\!)x^{14} \nonumber \\
&&\hspace{-7mm}\quad + \!({\displaystyle \frac {676039}{8712}} \mu
^{13} \!- {\displaystyle \frac {29393}{121}} \mu ^{11} \!+\!
{\displaystyle \frac {104975}{363}} \mu ^{9} \!- {\displaystyle
\frac {176800}{ 1089}} \mu ^{7} \!+\! {\displaystyle \frac
{5200}{121}} \mu ^{5} \! - {\displaystyle \frac {1664}{363}} \mu
^{3} \!+\! {\displaystyle \frac {128}{1089}} \mu ) x^{13} \nonumber \\
&&\hspace{-7mm}\quad  + \!( {\displaystyle \frac {29393}{121}} \mu
^{12} \!-\! {\displaystyle \frac {3069469}{4356}} \mu ^{10} \!+\!
{\displaystyle \frac {1651975}{2178}} \mu ^{8} \!-\!
{\displaystyle \frac {403000}{1089}} \mu ^{6} \!\!+\!
{\displaystyle \frac {86320}{1089}} \mu ^{4} \!-\! {\displaystyle
\frac {6368}{ 1089}}
\mu ^{2} \!\!+\! {\displaystyle \frac {64}{1089}})x^{12}  \nonumber \\
&&\hspace{-7mm}\quad + ({\displaystyle \frac {41990}{99}}  \mu ^{
11} - {\displaystyle \frac {224315}{198}}  \mu ^{9} + {\displaystyle
\frac {107900}{99}}  \mu ^{7} - {\displaystyle \frac {44720}{99}}
 \mu ^{5} + {\displaystyle \frac {7360}{99}}  \mu ^{3} -
{\displaystyle \frac {320}{99}}  \mu ) x^{11}
 \nonumber \\
 &&\hspace{-7mm}\quad    + ({\displaystyle \frac {44200}{99}}  \mu ^{10} -
{\displaystyle \frac {107900}{99}}  \mu ^{8} + {\displaystyle \frac
{91429}{99}}  \mu ^{6} - {\displaystyle \frac {31192}{99} }  \mu
^{4} + {\displaystyle \frac {3632}{99}}  \mu ^{2} -
{\displaystyle \frac {64}{99}} ) x^{10} \nonumber \\
 &&\hspace{-7mm}\quad  + ({\displaystyle \frac {3120}{11}}  \mu ^{9} -
{\displaystyle \frac {20696}{33}}  \mu ^{7} + {\displaystyle \frac
{15154}{33}}  \mu ^{5} - {\displaystyle \frac {4064}{33}}
 \mu ^{3} + {\displaystyle \frac {96}{11}}  \mu ) x^{9} \nonumber \\
 &&\hspace{-7mm}\quad    + ({\displaystyle \frac {3328}{33}}  \mu ^{8} -
{\displaystyle \frac {6592}{33}}  \mu ^{6} + {\displaystyle \frac
{4064}{33}}  \mu ^{4} - {\displaystyle \frac {796}{33}}   \mu
^{2} + {\displaystyle \frac {8}{11}} ) x^{8} \nonumber \\
&&\hspace{-7mm}\quad + ({\displaystyle \frac {512}{33}}  \mu ^{7} -
{\displaystyle \frac {896}{33}}  \mu ^{5} + {\displaystyle \frac
{448}{33}}  \mu ^{3} - {\displaystyle \frac {56}{33}}
\mu ) x^{7}\ , \label{eqb7} \\
&& \nonumber \\
&&\hspace{-7mm}\vf^{(8)}(x,\mu) = ( \!-\! {\displaystyle \frac
{1077205}{81796}}   \mu ^{16} \!+\! {\displaystyle \frac
{1040060}{20449}}  \mu ^{14} \!-\! {\displaystyle \frac
{14560840}{184041}}  \mu ^{12} \!+\! {\displaystyle \frac
{11648672}{184041}}  \mu ^{10} \!-\!
{\displaystyle \frac {5064640}{184041}}  \mu ^{8} \nonumber \\
 &&\hspace{-7mm}\quad    \!+\! {\displaystyle \frac {1157632}{184041}}  \mu ^{6}
 \!-\! {\displaystyle \frac {121856}{184041}}  \mu ^{4} \!+\!
{\displaystyle \frac {4096}{184041}}  \mu ^{2})x^{16}\mbox{} \!+\!
(
 \!-\! {\displaystyle \frac {2228700}{20449}}  \mu ^{15} \!+\!
{\displaystyle \frac {72804200}{184041}}  \mu ^{13} \nonumber \\
 &&\hspace{-7mm}\quad    \!- {\displaystyle \frac {11648672}{20449}}  \mu ^{11
} \!+\! {\displaystyle \frac {25323200}{61347}}  \mu ^{9} \!-\!
{\displaystyle \frac {28940800}{184041}}  \mu ^{7} \!+\!
{\displaystyle \frac {609280}{20449}}  \mu ^{5} \!-\!
{\displaystyle \frac {143360}{61347}}  \mu ^{3}  \nonumber \\
 &&\hspace{-7mm}\quad    \!+\! {\displaystyle \frac {8192}{184041}}  \mu )x^{15}
\mbox{} \!+\! ( \!-\! {\displaystyle \frac {72804200}{184041}} \mu
^{14 } \!+\! {\displaystyle \frac {247326268}{184041}}  \mu ^{12}
\!-\!
{\displaystyle \frac {36393056}{20449}}  \mu ^{10} \nonumber \\
 &&\hspace{-7mm}\quad    \!+\! {\displaystyle \frac {212663200}{184041}}  \mu ^{
8} \!-\! {\displaystyle \frac {69632000}{184041}}  \mu ^{6} \!+\!
{\displaystyle \frac {1169920}{20449}}  \mu ^{4} \!-\!
{\displaystyle \frac {561152}{184041}}  \mu ^{2} \!+\!
{\displaystyle \frac {4096}{184041}} )x^{14} \nonumber \\
&&\hspace{-7mm}\quad \!- {\displaystyle \frac {11648672}{14157}}
 \mu ^{13} \!+\! {\displaystyle \frac {36863344}{14157}}  \mu ^{11}
\!-\! {\displaystyle \frac {44548160}{14157}}  \mu ^{9} \!+\!
{\displaystyle \frac {25568000}{14157}}  \mu ^{7} \!-\!
{\displaystyle \frac {7014400}{14157}}  \mu ^{5} \nonumber \\
 &&\hspace{-7mm}\quad    +{\displaystyle \frac {800768}{14157}}  \mu ^{3}
 \!-\! {\displaystyle \frac {8192}{4719}}  \mu )x^{13}\mbox{} \!+\! ( \!-\!
{\displaystyle \frac {5064640}{4719}}  \mu ^{12} \!+\!
{\displaystyle \frac {44548160}{14157}}  \mu ^{10} \!-\!
{\displaystyle \frac {48665560}{14157}}  \mu ^{8} \nonumber \\
 &&\hspace{-7mm}\quad   \!+\! {\displaystyle \frac {24267200}{14157}}  \mu ^{6}
 \!-\! {\displaystyle \frac {5384960}{14157}}  \mu ^{4} \!+\!
{\displaystyle \frac {424960}{14157}}  \mu ^{2} \!-\!
{\displaystyle \frac {5120}{14157}} )x^{12} \nonumber \\
&&\hspace{-7mm}\quad ( \!-\! {\displaystyle \frac {1157632}{1287}}
 \mu ^{11} \!+\! {\displaystyle \frac {3124736}{1287}}  \mu ^{9} \!-\!
{\displaystyle \frac {3050176}{1287}}  \mu ^{7} \!+\!
{\displaystyle \frac {1291648}{1287}}  \mu ^{5} \!-\!
{\displaystyle \frac {220160}{1287}}  \mu ^{3} \!+\!
{\displaystyle
\frac {10240}{1287}}  \mu ) x^{11} \nonumber \\
 &&\hspace{-7mm}\quad \!+\! ( \!-\! {\displaystyle \frac {609280}{1287}}  \mu ^{
10} \!+\! {\displaystyle \frac {1502720}{1287}}  \mu ^{8} \!-\!
{\displaystyle \frac {1291648}{1287}}  \mu ^{6} \!+\!
{\displaystyle \frac {34624}{99}}  \mu ^{4} \!-\! {\displaystyle
\frac {54272}{1287}}
 \mu ^{2} \!+\! {\displaystyle \frac {1024}{
1287}} ) x^{10} \nonumber \\
 &&\hspace{-7mm}\quad    \!+\! ( \!-\! {\displaystyle \frac {20480}{143}}  \mu ^{9}
 \!+\! {\displaystyle \frac {137216}{429}}  \mu ^{7} \!-\!
{\displaystyle \frac {101888}{429}}  \mu ^{5} \!+\! {\displaystyle
\frac {27904}{429}}  \mu ^{3} \!-\! {\displaystyle \frac
{2048}{429}}  \mu ) x^{9} \nonumber \\
 &&\hspace{-7mm}\quad    \!+\! ( \!-\! {\displaystyle \frac {8192}{429}}  \mu ^{8}
 \!+\! {\displaystyle \frac {16384}{429}}  \mu ^{6} \!-\!
{\displaystyle \frac {10240}{429}}  \mu ^{4} \!+\! {\displaystyle
\frac {2048}{429}}  \mu ^{2} \!-\! {\displaystyle \frac {64}{429}}
) x^{8} \ , \label{eqb8} \\
&& \nonumber \\
&&\hspace{-7mm}\vf^{(9)}(x,\mu) = ({\displaystyle \frac
{1178589}{74360}}  \mu ^{18} \!-\! {\displaystyle \frac
{7071534}{102245}}  \mu ^{16} \!+\! {\displaystyle \frac
{12774384}{102245}}  \mu ^{14} \!-\! {\displaystyle \frac
{12333888}{102245}}  \mu ^{12} \!+\!
{\displaystyle \frac {1370432}{20449}}  \mu ^{10} \nonumber \\
 &&\hspace{-7mm}\quad    \!- {\displaystyle \frac {10963456}{511225}}  \mu ^{8
} \!+\! {\displaystyle \frac {1906688}{511225}}  \mu ^{6} \!-\!
{\displaystyle \frac {155648}{511225}}  \mu ^{4} \!+\!
{\displaystyle \frac {4096}{511225}}  \mu ^{2})x^{18}\mbox{} \!+\!
({\displaystyle \frac {60108039}{408980}}  \mu ^{17} \nonumber \\
&&\hspace{-7mm}\quad \!- {\displaystyle \frac {62047008}{102245}}
\mu ^{ 15} \!+\! {\displaystyle \frac {104838048}{102245}}  \mu
^{13} \!-\! {\displaystyle \frac {93189376}{102245}}  \mu ^{11}
\!+\! {\displaystyle \frac {46594688}{102245}}  \mu ^{9} \!-\!
{\displaystyle \frac {64827392}{511225}}  \mu ^{7} \nonumber \\
 &&\hspace{-7mm}\quad   \!+\! {\displaystyle \frac {9261056}{511225}}  \mu ^{5}
 \!-\! {\displaystyle \frac {557056}{511225}}  \mu ^{3} \!+\!
{\displaystyle \frac {8192}{511225}}  \mu )x^{17}\mbox{} \!+\! (
{\displaystyle \frac {62047008}{102245}}  \mu ^{16} \!-\!
{\displaystyle \frac {18587358}{7865}}  \mu ^{14} \nonumber \\
&&\hspace{-7mm}\quad \!+\! {\displaystyle \frac
{380037924}{102245}}  \mu ^{ 12} \!-\! {\displaystyle \frac
{308689808}{102245}}  \mu ^{10} \!+\! {\displaystyle \frac
{137251744}{102245}}  \mu ^{8} \!-\!
{\displaystyle \frac {162647296}{511225}}  \mu ^{6} \nonumber \\
&&\hspace{-7mm}\quad \!+\! {\displaystyle \frac
{18339328}{511225}}  \mu ^{4 } \!-\! {\displaystyle \frac
{739328}{511225}}  \mu ^{2} \!+\! {\displaystyle \frac
{4096}{511225}} )x^{16}\mbox{} \!+\! ( {\displaystyle \frac
{29953728}{20449}}  \mu ^{15} \!-\!
{\displaystyle \frac {109622324}{20449}}  \mu ^{13} \nonumber \\
&&\hspace{-7mm}\quad  \!+\! {\displaystyle \frac
{797934032}{102245}}  \mu ^{ 11} \!-\! {\displaystyle \frac
{586991776}{102245}}  \mu ^{9} \!+\! {\displaystyle \frac
{45726464}{20449}}  \mu ^{7} \!-\! {\displaystyle \frac
{4082176}{9295}}  \mu ^{5} \!+\!
{\displaystyle \frac {3770368}{102245}}  \mu ^{3} \nonumber \\
 &&\hspace{-7mm}\quad  \!- {\displaystyle \frac {86016}{102245}}  \mu )x^{15
}\mbox{} \!+\! ({\displaystyle \frac {46594688}{20449}}  \mu ^{14}
 \!-\! {\displaystyle \frac {797934032}{102245}}  \mu ^{12} \!+\!
{\displaystyle \frac {1068177796}{102245}}  \mu ^{10} \nonumber \\
&&\hspace{-7mm}\quad \!- {\displaystyle \frac {703969456}{102245}}
\mu ^{ 8} \!+\! {\displaystyle \frac {47124544}{20449}}  \mu ^{6}
\!-\! {\displaystyle \frac {36921856}{102245}}  \mu ^{4} \!+\!
{\displaystyle \frac {2105856}{102245}}  \mu ^{2} \!-\!
{\displaystyle \frac {18432}{102245}} )x^{14}\nonumber \\
&&\hspace{-7mm}\quad \!+\!{\displaystyle \frac {93189376}{39325}}
 \mu ^{13} \!-\! {\displaystyle \frac {297294368}{39325}}  \mu
^{11} \!+\! {\displaystyle \frac {363121768}{39325}}  \mu ^{9}
\!-\! {\displaystyle \frac {16269952}{3025}}  \mu ^{7} \!+\!
{\displaystyle \frac {11860864}{7865}}  \mu ^{5} \nonumber \\
&&\hspace{-7mm}\quad \!- {\displaystyle \frac {280576}{1573}}
 \mu ^{3} \!+\! {\displaystyle \frac {9216}{1573}}  \mu
)x^{13}\mbox{} \!+\! ( {\displaystyle \frac {64827392}{39325}}
 \mu ^{12} \!-\! {\displaystyle \frac {191588096}{39325}}  \mu
^{10} \!+\!{\displaystyle \frac {16269952}{3025}}  \mu ^{8}  \nonumber \\
 &&\hspace{-7mm}\quad  \!- {\displaystyle \frac {107020512}{39325}}  \mu ^{6
} \!+\! {\displaystyle \frac {970688}{1573}}  \mu ^{4} \!-\!
{\displaystyle \frac {79360}{1573}}  \mu ^{2} \!+\! {\displaystyle
\frac {1024}{1573}} )x^{12} \nonumber \\
 &&\hspace{-7mm}\quad    \!+\! ({\displaystyle \frac {2646016}{3575}}  \mu ^{11} \!-\!
{\displaystyle \frac {7198208}{3575}}  \mu ^{9} \!+\!
{\displaystyle \frac {7099648}{3575}}  \mu ^{7} \!-\!
{\displaystyle \frac {3050176}{3575}}  \mu ^{5} \!+\!
{\displaystyle \frac {21248}{143}}  \mu ^{3} \!-\! {\displaystyle
\frac {1024}{143}}  \mu ) x^{11} \nonumber \\
 &&\hspace{-7mm}\quad    \!+\! ({\displaystyle \frac {139264}{715}}  \mu ^{10}
 \!-\! {\displaystyle \frac {26624}{55}}  \mu ^{8} \!+\! {\displaystyle
\frac {300544}{715}}  \mu ^{6} \!-\! {\displaystyle \frac {21248}{
143}}  \mu ^{4} \!+\! {\displaystyle \frac {13088}{715}}  \mu ^{2}
 \!-\! {\displaystyle \frac {256}{715}} ) x^{10} \nonumber \\
&&\hspace{-7mm}\quad \!+\! ({\displaystyle \frac {16384}{715}}
 \mu ^{9} \!-\! {\displaystyle \frac {36864}{715}}  \mu ^{7}
\!+\! {\displaystyle \frac {27648}{715}}  \mu ^{5} \!-\!
{\displaystyle \frac {1536}{143} }  \mu ^{3} \!+\! {\displaystyle
\frac {576}{715}}  \mu ) x^{9} \ .  \label{eqb9}\eey


\vv\vv

\begin{center}
{\bf APPENDIX C: EXPRESSIONS FOR THE EVEN MULTIPOLE SOLUTIONS OF
ORDERS $n = 0$ TO $n = 9$ IN THE \\ CYLINDRICAL COORDINATE SYSTEM}
\end{center}

\setcounter{equation}{-1}
\def\theequation{C.\arabic{equation}}

\vv

For simplicity of notation the variables are chosen to be:
\[
\xi \equiv \left(\frac{R}{R_A}\right)^2\ , \ \ \ \nu \equiv
\left(\frac{z}{R_A}\right)^2 \ ,
\]
where $R$ and $z$ are the cylindrical coordinates of a point and
$R_A$ is the radial coordinate defining the circle on the equatorial
plane where the gradients of the multipole solutions are designated
to be zero. \bey \vf^{(0)}(\xi ,
\nu ) &=& 1\ , \label{eqc0} \\
&& \nonumber \\
\vf^{(1)}(\xi ,  \nu ) &=&  {\displaystyle \frac {1}{4}}(\xi - 1)\
, \label{eqc1} \\
&& \nonumber \\
\vf^{(2)}(\xi ,  \nu ) &=&  - {\displaystyle \frac {(\xi  - 1)^{2}}{
4}} + \nu  (\xi  - 1) + \nu\ , \label{eqc2} \\
&& \nonumber \\
\vf^{(3)}(\xi ,  \nu ) &=& {\displaystyle \frac {(\xi  - 1)^{3}}{8}
} - {\displaystyle \frac {3 \nu  (\xi  - 1)^{2}}{2}}  + (\nu ^{2} -
{\displaystyle \frac {3}{2}}  \nu ) (\xi  - 1) + \nu ^{2 } \ ,
\label{eqc3} \\
&& \nonumber \\
\vf^{(4)}(\xi ,  \nu ) &=&  - {\displaystyle \frac {(\xi - 1)^{4}}{
20}} + {\displaystyle \frac {6 \nu  (\xi  - 1)^{3}}{5}} + (
 - {\displaystyle \frac {12}{5}}  \nu ^{2} + {\displaystyle
\frac {6}{5}}  \nu ) (\xi  - 1)^{2} \nonumber \\
&&+ ({\displaystyle \frac {16 }{25}}  \nu ^{3} - {\displaystyle
\frac {16}{5}}  \nu ^{2}) ( \xi  - 1) - {\displaystyle \frac {4 \nu
^{2}}{5}}  + {\displaystyle
\frac {16 \nu ^{3}}{25}} \ , \label{eqc4} \\
&& \nonumber \\
\vf^{(5)}(\xi ,  \nu ) &=& {\displaystyle \frac {(\xi - 1)^{5}}{56}
} - {\displaystyle \frac {5 \nu  (\xi  - 1)^{4}}{7}} + (
{\displaystyle \frac {20}{7}}  \nu ^{2} - {\displaystyle \frac {
5}{7}}  \nu ) (\xi - 1)^{3} \nonumber \\
&&+ ( - {\displaystyle \frac {16}{7} }
 \nu ^{3} + {\displaystyle \frac {30}{7}}  \nu ^{2}) (\xi
 - 1)^{2} \nonumber \\
&& + ({\displaystyle \frac {16}{49}}  \nu ^{4} - {\displaystyle
\frac {24}{7}}  \nu ^{3} + {\displaystyle \frac { 10}{7}}  \nu ^{2})
(\xi  - 1) + {\displaystyle \frac {16 \nu ^{4}}{49}}  -
{\displaystyle \frac {8 \nu ^{3}}{7}}  \ , \label{eqc5} \\
&& \nonumber \\
\vf^{(6)}(\xi ,  \nu ) &=&  - {\displaystyle \frac {(\xi - 1)^{6}}{
168}} + {\displaystyle \frac {5 \nu  (\xi  - 1)^{5}}{14}} + (
 - {\displaystyle \frac {50}{21}}  \nu ^{2} + {\displaystyle
\frac {5}{14}}  \nu ) (\xi  - 1)^{4} + ({\displaystyle \frac {
80}{21}}  \nu ^{3} \nonumber \\
&&- {\displaystyle \frac {80}{21}}  \nu ^{2})
 (\xi  - 1)^{3} + ( - {\displaystyle \frac {80}{49}}  \nu ^{4} + {\displaystyle
\frac {48}{7}}  \nu ^{3} - {\displaystyle \frac { 10}{7}}  \nu ^{2})
(\xi  - 1)^{2} \nonumber \\
&&+ ({\displaystyle \frac {64}{ 441}}  \nu ^{5} - {\displaystyle
\frac {128}{49}}  \nu ^{4} + {\displaystyle \frac {24}{7}}  \nu
^{3}) (\xi  - 1) - {\displaystyle \frac {48 \nu ^{4}}{49}}  +
{\displaystyle \frac {8 \nu ^{3}}{21}} + {\displaystyle \frac
{64 \nu ^{5}}{441}} \ ,  \label{eqc6} \\
&& \nonumber \\
 \vf^{(7)}(\xi ,  \nu ) &=& {\displaystyle \frac
{(\xi - 1)^{7}}{528 }} - {\displaystyle \frac {7 \nu  (\xi  -
1)^{6}}{44}} + ( {\displaystyle \frac {35}{22}}  \nu ^{2} -
{\displaystyle \frac {7}{44}}  \nu ) (\xi  - 1)^{5} \nonumber \\
&& + ( - {\displaystyle \frac {140}{33}}
 \nu ^{3} + {\displaystyle \frac {175}{66}}
\nu ^{2}) (\xi  - 1)^{4} + ({\displaystyle \frac {40}{11}} \nu ^{4}
- {\displaystyle \frac {280}{33}}  \nu ^{3} + {\displaystyle \frac
{35}{33}}  \nu ^{2}) (\xi  - 1)^{3} \nonumber \\
&&+ ( - {\displaystyle \frac {32}{33}}  \nu ^{5} + {\displaystyle
\frac {80}{11}}  \nu
 ^{4} - {\displaystyle \frac {56}{11}}  \nu ^{3}) (\xi  - 1)^{2
} \nonumber \\
&&+ ({\displaystyle \frac {64}{1089}}  \nu ^{6} - {\displaystyle
\frac {160}{99}}  \nu ^{5} + {\displaystyle \frac {48}{11}}  \nu
^{4} - {\displaystyle \frac {28}{33}}  \nu
 ^{3}) (\xi  - 1) \nonumber \\
 && + {\displaystyle \frac {8 \nu ^{4}}{11}}  +
{\displaystyle \frac {64 \nu ^{6}}{1089}}  - {\displaystyle \frac
{64 \nu ^{5}}{99}}  \ , \label{eqc7} \\
&& \nonumber \\
\vf^{(8)}(\xi ,  \nu ) &=&  - {\displaystyle \frac {(\xi - 1)^{8}}{
1716}}  + {\displaystyle \frac {28 \nu  (\xi  - 1)^{7}}{429}}
 + ( - {\displaystyle \frac {392}{429}}  \nu ^{2} +
{\displaystyle \frac {28}{429}}  \nu ) (\xi  - 1)^{6} \nonumber \\
&& + ({\displaystyle \frac {1568}{429}}  \nu ^{3} - {\displaystyle
\frac {224}{143}}  \nu ^{2}) (\xi  - 1)^{5} + (
 - {\displaystyle \frac {2240}{429}}  \nu ^{4} + {\displaystyle
\frac {1120}{143}}  \nu ^{3} - {\displaystyle \frac {280}{429}}
 \nu ^{2}) (\xi  - 1)^{4} \nonumber \\
&& + ({\displaystyle \frac {3584}{1287}}  \nu ^{5} - {\displaystyle
\frac {5120}{429}}  \nu ^{4} + {\displaystyle
\frac {2240}{429}}  \nu ^{3}) (\xi  - 1)^{3} \nonumber \\
&& + ( - {\displaystyle \frac {7168}{14157}}  \nu ^{6} +
{\displaystyle \frac {2560}{429}}  \nu ^{5} - {\displaystyle \frac
{1280}{143}}  \nu ^{4} + {\displaystyle \frac {448}{429}}
 \nu ^{3}) (\xi  - 1)^{2} \nonumber \\
&& + ({\displaystyle \frac {4096}{184041}}  \nu ^{7} -
{\displaystyle \frac {4096}{4719}}  \nu ^{6} + {\displaystyle \frac
{5120}{1287}}  \nu ^{5} - {\displaystyle \frac {1024}{429} }  \nu
^{4}) (\xi  - 1) \nonumber \\
&&+ {\displaystyle \frac {4096 \nu ^{7} }{184041}}  - {\displaystyle
\frac {5120 \nu ^{6}}{14157}} + {\displaystyle \frac {1024 \nu
^{5}}{1287}}  - {\displaystyle
\frac {64 \nu ^{4}}{429}}  \ , \label{eqc8} \\
&& \nonumber \\
\vf^{(9)}(\xi ,  \nu ) &=& {\displaystyle \frac {(\xi - 1)^{9}}{
5720}} - {\displaystyle \frac {18 \nu  (\xi  - 1)^{8}}{715}}
 + ({\displaystyle \frac {336}{715}}  \nu ^{2} - {\displaystyle
\frac {18}{715}}  \nu ) (\xi  - 1)^{7} \nonumber \\
&& + ( - {\displaystyle \frac {9408}{3575}}  \nu ^{3} +
{\displaystyle \frac {588}{715}}  \nu ^{2}) (\xi  - 1)^{6} \nonumber
\\
&&+ ( {\displaystyle \frac {4032}{715}}  \nu ^{4} - {\displaystyle
\frac {21168}{3575}}  \nu ^{3} + {\displaystyle \frac {252}{715}
}  \nu ^{2}) (\xi  - 1)^{5} \nonumber \\
&& + ( - {\displaystyle \frac {3584}{715}}  \nu ^{5} +
{\displaystyle \frac {2016}{143}}  \nu ^{4} - {\displaystyle
\frac {3024}{715}}  \nu ^{3}) (\xi  - 1)^{4} \nonumber \\
&& + ({\displaystyle \frac {14336}{7865}}  \nu ^{6} - {\displaystyle
\frac {1792}{143}}  \nu ^{5} + {\displaystyle \frac {1728}{143}}
\nu ^{4} - {\displaystyle \frac {672}{715}}
 \nu ^{3}) (\xi  - 1)^{3} \nonumber \\
&& + ( - {\displaystyle \frac {24576}{102245}}  \nu ^{7} +
{\displaystyle \frac {32256}{7865}}  \nu ^{6} - {\displaystyle \frac
{1536}{143}}  \nu ^{5} + {\displaystyle \frac {576}{143}}
 \nu ^{4}) (\xi  - 1)^{2} \nonumber \\
&& + ({\displaystyle \frac {4096}{511225}}  \nu ^{8} -
{\displaystyle \frac {43008}{102245}}  \nu ^{7} + {\displaystyle
\frac {4608}{1573}}  \nu ^{6} - {\displaystyle \frac {512}{143}}
 \nu ^{5} + {\displaystyle \frac {288}{715}}  \nu ^{4}) (\xi  -
1) \nonumber \\
&&- {\displaystyle \frac {18432 \nu ^{7}}{ 102245}}  +
{\displaystyle \frac {1024 \nu ^{6}}{1573}}  - {\displaystyle \frac
{256 \nu ^{5}}{715}}  + {\displaystyle \frac {4096 \nu
^{8}}{511225}}  \ . \label{eqc9} \eey

\vv

\begin{center}

\textbf{DEDICATION}

\end{center}

\vv

This paper is dedicated to the memory of Professor Gumercindo Lima
who, as professor of the author in high school, taught him the
relations in combinatorial analysis that are applied in the theory
here presented.

\end{document}